\documentclass[twocolumn]{revtex4}
\usepackage[intlimits]{amsmath}
\usepackage{amssymb}
\usepackage{graphicx}
\usepackage{epstopdf}
\usepackage{hyperref}

\usepackage{natbib}
\begin{document}
\title{Atomistics of vapor-liquid-solid nanowire growth}

\author{Hailong Wang}
\affiliation{Group for Simulation and Theory of Atomic-Scale Material Phenomena ({\it st}AMP), Department of Mechanical and Industrial Engineering, Northeastern University, Boston MA 02115}
\affiliation{School of Engineering, Brown University, Providence, Rhode Island 02912, USA}
\author{Luis A. Zepeda-Ruiz}
\affiliation{Physical and Life Sciences Directorate, Lawrence Livermore National Laboratory, Livermore, CA 94550}
\author{George H. Gilmer}
\affiliation{Physical and Life Sciences Directorate, Lawrence Livermore National Laboratory, Livermore, CA 94550}
\affiliation{Division of Engineering, Colorado School of Mines, Golden, CO 80401}
\author{Moneesh Upmanyu}
\affiliation{Group for Simulation and Theory of Atomic-Scale Material Phenomena ({\it st}AMP), Department of Mechanical and Industrial Engineering, Northeastern University, Boston MA 02115}
\affiliation{Department of Bioengineering, Northeastern University, Boston MA 02115}
\email{mupmanyu@neu.edu}

\begin{abstract}
Vapor-liquid-solid (VLS) route and its variants are routinely used for scalable synthesis of semiconducting nanowires yet the fundamental processes remain unknown. Here, we employ atomic-scale computations based on model potentials to study the stability and growth of gold-catalyzed silicon nanowires (SiNWs). Equilibrium studies uncover segregation at the solid-like surface of the  catalyst particle, a liquid AuSi droplet, and a silicon-rich droplet-nanowire interface enveloped by heterogeneous truncating facets. Supersaturation of the droplets leads to rapid 1D growth on the truncating facets and much slower nucleation-controlled 2D growth on the main facet. Surface diffusion is suppressed and the excess Si flux occurs through the droplet bulk which, together with the Si-rich interface and contact line, lowers the nucleation barrier on the main facet. The ensuing step flow is modified by Au diffusion away from the step edges. Our study highlights key interfacial characteristics for morphological and compositional control of semiconducting nanowire arrays.
\end{abstract}
\maketitle

\section*{Introduction}
\begin{figure*}
\centering
\includegraphics[width=2\columnwidth]{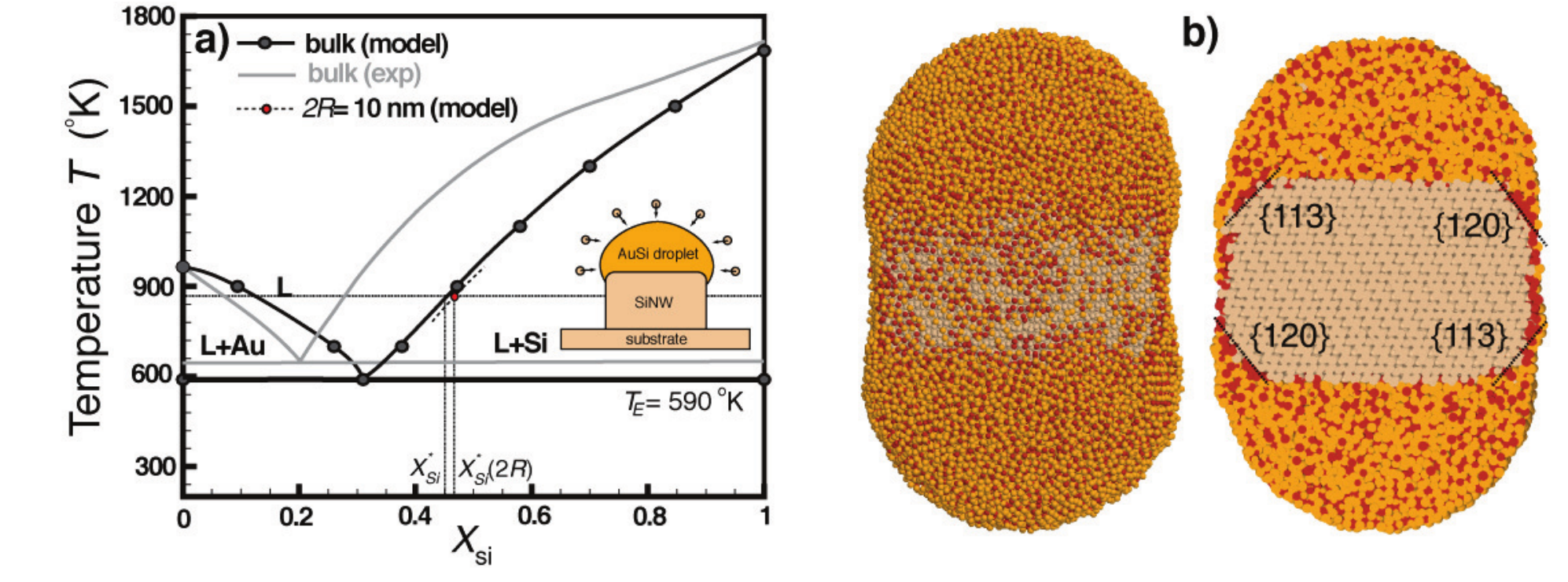}
\caption{{\bf Phase diagram of the model Au-Si  system}: (a) Bulk phase diagram for the Au-Si binary system predicted by the model inter-atomic interactions employed in this study (solid black line) and the size correction to the model diagram $X_{Si}^{\ast}(2R)=0.46$ (solid red circle) based on doubly-capped SiNW computations shown in (b). The solid gray line is the experimental bulk phase diagram shown for comparison. 
(inset) Schematic illustrating the VLS route for SiNW nanowire synthesis. (b) (left) Atomic configuration of an equilibrated 10\,nm long SiNW capped with catalyst particles at both ends. (right) Mid-section slice through the droplet-nanowire system that reveals the shape of the droplets and the faceted nature of the solid nanowire.  Light yellow is solid silicon, and red and bright yellow indicate liquid Si and Au, respectively. For consistency, the color scheme for the atomic plots and the density profiles is preserved in the remaining figures. 
\label{fig:figure1}
}
\end{figure*}

A characteristic feature of the VLS route for nanowire synthesis is the presence of a catalyst particle, usually a droplet, that mediates the mass transfer from the vapor phase precursor to the growing nanowire\cite{nw:WagnerEllis:1964, nw:GivargizovChernov:1973, nw:WuLieber:2004, nw:SchmidtGosele:2005, nw:SchmidtGosele:2009, nw:Ross:2010}. The technique was demonstrated successfully in the 1960s in the context of crystalline whisker growth\cite{nw:WagnerEllis:1964}, and has become increasingly relevant of late as it offers direct control over nanowire composition and morphology, aspects of critical importance in several nanowire-based applications\cite{nw:Lieber:1998, nw:Lieber:2001, nw:Lieber:2003, nano:BarkerArias:2002, nw:SamuelsonBjork:2004, nw:RoperVoorhees:2007, nw:RossTersoffReuter:2005, nw:SchmidtGosele:2007, Thompson2009, nw:SchwalbachVoorhees:2009, nw:MadrasDrucker:2009, nw:IrreraPecoraPriolo:2009}. The salient features can be readily seen in the schematic illustration in  Fig.~\ref{fig:figure1}a (inset). A low melting eutectic droplet  is supersaturated following breakdown of the precursor gas on its surface. Past a critical point, crystallization at the droplet-nanowire interface results in tip growth. The droplet then acts as a conduit for 1D crystal growth with diameters that are set by its dimensions.

The atomic-scale structure, energetics and dynamics of the alloyed particle directly influence the morphology and composition of the as-grown nanowires, yet the prevailing ideas continue to be based on continuum frameworks\cite{nw:Givargizov:1975, nw:DubrovskiNickolai:2004, nw:SchmidtGosele:2007,nw:RoperVoorhees:2007,nw:SchwarzTersoff:2009, nw:DubrovskiiGlas:2009, nw:SchwarzTersoff:2011, nw:SchwarzTersoff:2011b}.  As an example, gold-catalyzed growth of silicon nanowires is a well-studied system yet the morphology and composition of the particle-nanowire interface is unknown. 
Basic crystal growth parameters such as the size dependent equilibrium composition $X_{Si}^{\ast}$ above which the droplet rejects the excess silicon onto the nanowire remain unknown; continuum models models ignore the effect of segregation and related interfacial phenomena. Such microscopic features modify the droplet morphology through the balance of interfacial tension at the contact line which in turn dynamically regulates the chemical potential difference between the two phases, i.e. the driving force for growth. Likewise, the atomic-scale processes that sustain the growth of the nanowire remain unknown.  
The steady-state is limited by reduction of the precursor gas on the droplet surface\cite{nw:KodambakaRoss:2006} and the 2D nucleation-limited step flow at the droplet-nanowire interface\cite{nw: KimRoss:2008, nw:KimRoss:2009}. Understanding the interplay with the dynamics that feeds the growth of the nuclei is crucial for compositional control of the as-grown nanowire\cite{nw:PereaLauhon:2006,nw:PereaLauhon:2009,nw:SchwalbachVoorhees:2009, nw:HemesathVoorheesLauhon:2012} and requires an atomistic understanding of the near-equilibrium particle behavior.

We focus exclusively on a small diameter $2R=10$\,nm, $\langle111\rangle$ silicon nanowire growing isothermally above the eutectic temperature at $T=873$\,K.
Both experiments and {\it ab-initio} computational techniques are handicapped in accessing the spatio-temporal scales associated with the energetics and 3D dynamics during growth\cite{nw:LeeHwang:2010, nw:HaxhimaliAstaHoyt:2009, nw:RyuCai:2011}. We capture these processes in their full complexity by employing atomistic simulations based on an angular embedded-atom-method (AEAM) framework for describing the  hybrid metallic-covalent interactions\cite{nw:DongareZhigilei:2009}.  The computed surfaces of equilibrated isolated droplets are ordered and Si-rich, yet decorated by a submonolayer of Au. The surface structure leads to sluggish atomic mobilities such that most of the excess Si flux during growth is directed through the droplet bulk. Studies on droplet-capped SiNWs show that the solid-liquid interface is again Si-rich and the nanowire sidewalls are decorated by Au monolayers. Of note is the interface morphology composed of a main facet that truncates into smaller staircase-like, stepped facets at the contact line. The latter grow rapidly in a 1D fashion while crystallization on the main facet proceeds via nucleation of 2D islands which then grow via step flow. Interestingly, the motion of the steps is limited by Au diffusion into the droplets and/or the nanowire sidewalls. Studies with varying levels of droplet supersaturation yield the nucleation barrier which is lower than that for homogeneously nucleation on a planar Si(111)-AuSi interface, highlighting the combined role of segregation, atomic diffusivities, and interface morphology in regulating nanowire growth. 

\section*{Results}
\subsection{Size corrected liquidus composition}
We first quantify the equilibrium state of the alloy system by extracting the liquidus composition, $X_{Si}^\ast(2R, T=873$\,K). We capture the effect of the solid-liquid interface and the contact line by extracting the equilibrium state of a  $2R=10$\,nm diameter, $\langle111\rangle$ SiNW capped by a droplet at each end. The geometry allows the nanowire to act as a source/sink as the droplets adjust to the size-and morphology-dependent liquidus composition. The starting configuration consists of a nanowire with a hexagonal cross-section bounded by $\{112\}$-like sidewall facets (Methods). In particular, we have performed separate computations for SiNWs with $\{110\}$ and \{112\} sidewall facets. The choices are motivated by past experiments that have shown that the sidewalls can be a mixture of $\{$112$\}$ and $\{110\}$ facets\cite{nw:RossTersoffReuter:2005,nw:DavidGentile:2008}. Figure~\ref{fig:figure1}b shows the atomic configuration of the equilibrated SiNW system, in this particular case bounded by $\{110\}$ facets. We immediately see the formation of an Au submonolayer on the sidewalls, consistent with prior experimental observations\cite{nw:HannonRossTromp:2006}.  
The cross-section maintains its six-fold symmetry for both $\{$110$\}$ or $\{$112$\}$ family of facets implying that the two surfaces are energetically quite close; we do not observe their co-existence\cite{nw:RossTersoffReuter:2005, nw:DavidGentile:2008}. This is not surprising as the thermodynamic shape is inherently size dependent\cite{gbe:Herring:1951} and the diameters of the synthesized SiNWs are much larger. The $\{112\}$ facet is unstable for Au coverages typically observed during nanowire growth\cite{nw:RossTersoffReuter:2005}, yet we do not observe axial serrations due to the small nanowire lengths. We also cannot rule out the effects of surface stresses at these small sizes\cite{nw:LiangUpmanyuHuang:2005} as they can change the relative stability between $\{$110$\}$ or $\{$112$\}$ family of facets. The average composition within the interior of the capping droplets is $X_{Si}^\ast (2R)=0.46$ and is indicated on the phase diagram in Fig.~\ref{fig:figure1}a.  The size-corrected supersaturation with respect to the bulk liquidus composition is $\Delta_X=X_{Si}^\ast(2R) - X_{Si}^\ast\approx1\%$ and the equivalent undercooling $\Delta_T\approx 18$\,K.

\subsection{Surface segregation, ordering and kinetics}
\begin{figure}
\centering
\includegraphics[width=\columnwidth]{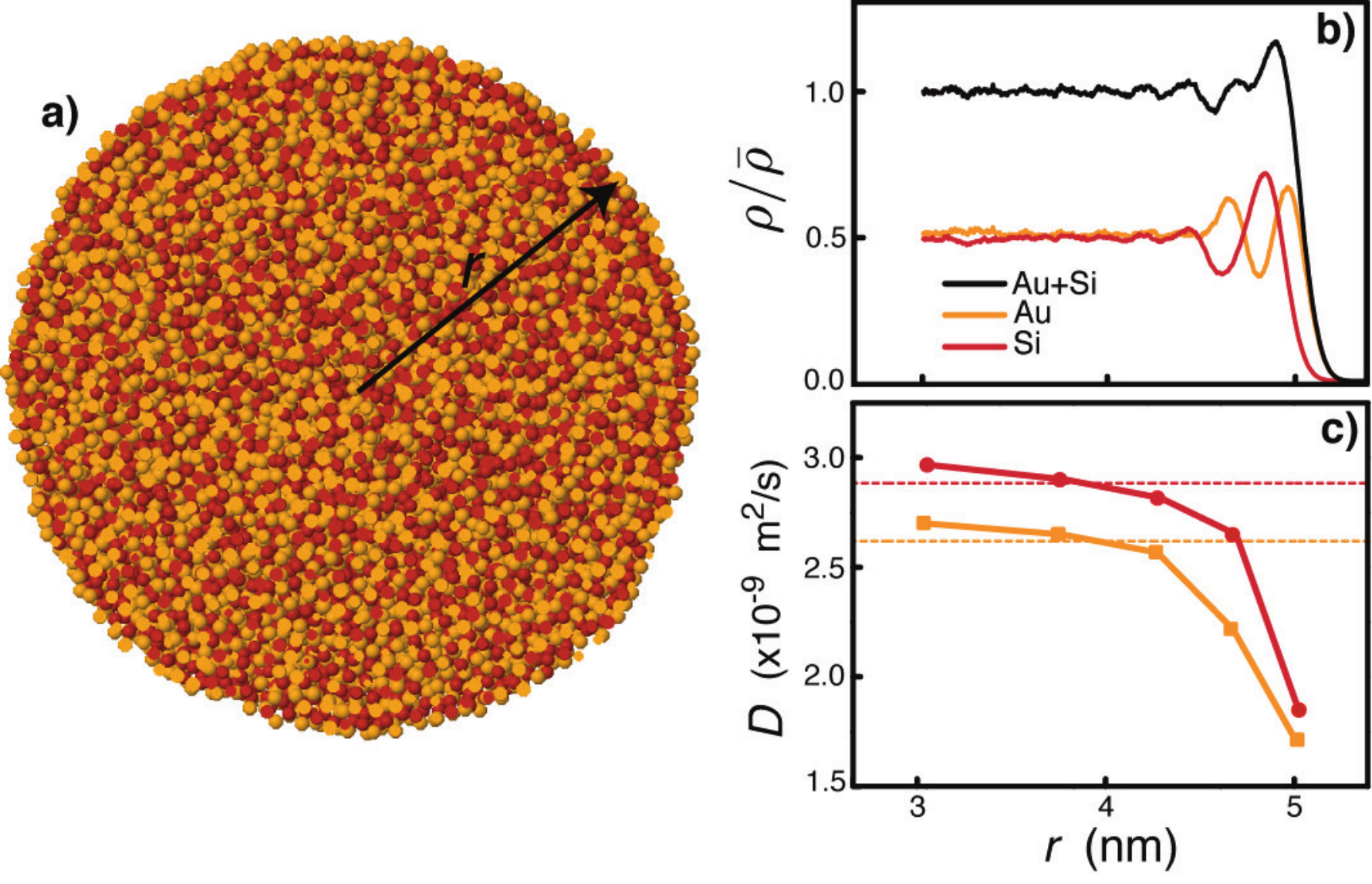}
\caption{\footnotesize {\bf Surface properties of equilibrated AuSi droplets}: (a) Midsection through an atomic plot  of a 32,000-atom AuSi system equilibrated at the model liquidus composition $X_{Si}^{\ast}=0.46$. 
(b and c) Radial variation of (b) the ensemble averaged and normalized density profile ${\rho}(r)/\bar{\rho}$  for the combined system (solid dark line) and for Au and Si, as indicated, and (c) ensemble averaged diffusivities of Au and Si, $D_{Au/Si}(r)$ extracted from equilibrium MD simulations of the droplets.  
The dotted lines in (c) correspond to the bulk diffusivities extracted from simulations on an equilibrated bulk alloy system (see Methods). 
\label{fig:figure2}
}
\end{figure}

The composition of the capping droplets is non-uniform and we decouple the surface effects by first studying isolated droplets at $X_{Si}=X_{Si}^\ast (2R)$. Atomic configuration of a midsection through an equilibrated droplet is shown in Fig.~\ref{fig:figure2}a. The non-uniform composition arises mainly due to surface segregation. Details are extracted using normalized density profiles $\langle\rho({\bf r})/\bar{\rho}\rangle$ plotted as a function of the distance $r$ from the droplet center of mass (Fig.~\ref{fig:figure2}b). The overall profile (dark solid line) has a well-defined peak at the surface suggestive of a solid-like surface structure. The multiple alternating peaks in the Au and Si profiles indicate a Si-rich subsurface  decorated by a submonolayer of Au. While the Si segregation is consistent with experiments on eutectic AuSi thin films\cite{tsf:ShpyrkoPershan:2006}, the Au submonolayer has not been reported earlier. 

We further isolate the effects of surface curvature via segregation studies on thin AuSi films and the results are summarized in Supplementary Figure~S1.  At the eutectic point, $X_{Si}\approx0.31$ and $T=590$\,K, we see ripples in the density profiles that extend to $\sim$7 layers and decay rapidly into the bulk. The density variations are larger than those usually seen on liquid surfaces and suggest a compositionally ordered pre-frozen surface qualitatively similar to the surface ordering seen in eutectic AuSi thin films\cite{tsf:ShpyrkoPershan:2006}. The ordering decreases with increasing $X_{Si}$ and temperature, and the extracted density profiles are similar to those observed in the liquid droplets, i.e. the surface curvature has little effect. The surfaces are always decorated by a Au submonolayer. It is expectedly less ordered compared to monolayers on crystalline Si surfaces, and this feature likely makes it difficult to be resolved by X-ray based surface characterization techniques\cite{tsf:ShpyrkoPershan:2006}. The discrepancy can certainly be an artifact of the EAM framework used to describe the metallic bonds\cite{intpot:WebbGrest:1986}; it underestimates the surface tension of liquid Au by 20\% and can therefore artificially favor the formation of the Au submonolayer. Nonetheless, it is natural to expect that the stability of Au monolayers on crystalline Si surfaces\cite{tsf:Lelay:1981} also extends to partially crystalline Si surfaces. As confirmation, we still see the Au decoration in segregations studies with modified Au-Au interactions using a charge gradient approach fit to the surface tension of liquid Au\cite{intpot:WebbGrest:1986} (Supplementary Figure~S2 and Supplementary Methods). The Au coverage is slightly reduced in extent yet the qualitative trends remain unchanged.
The effect of surface structure extends to the diffusive kinetics as well.  We quantify the interplay by monitoring the atomic mean-square displacements (MSD) that yield the atomic diffusivities $D_{Au/Si}$ (Supplementary Figure S3). The radial variation is plotted in Fig.~\ref{fig:figure2}c (Methods).  For both Au and Si, the surface diffusivity is less than half of the bulk value and further highlights the dramatic effect of the surface structure.

\subsection{Equilibrated nanowire-particle system}
\begin{figure*}
\centering
\includegraphics[width=2\columnwidth]{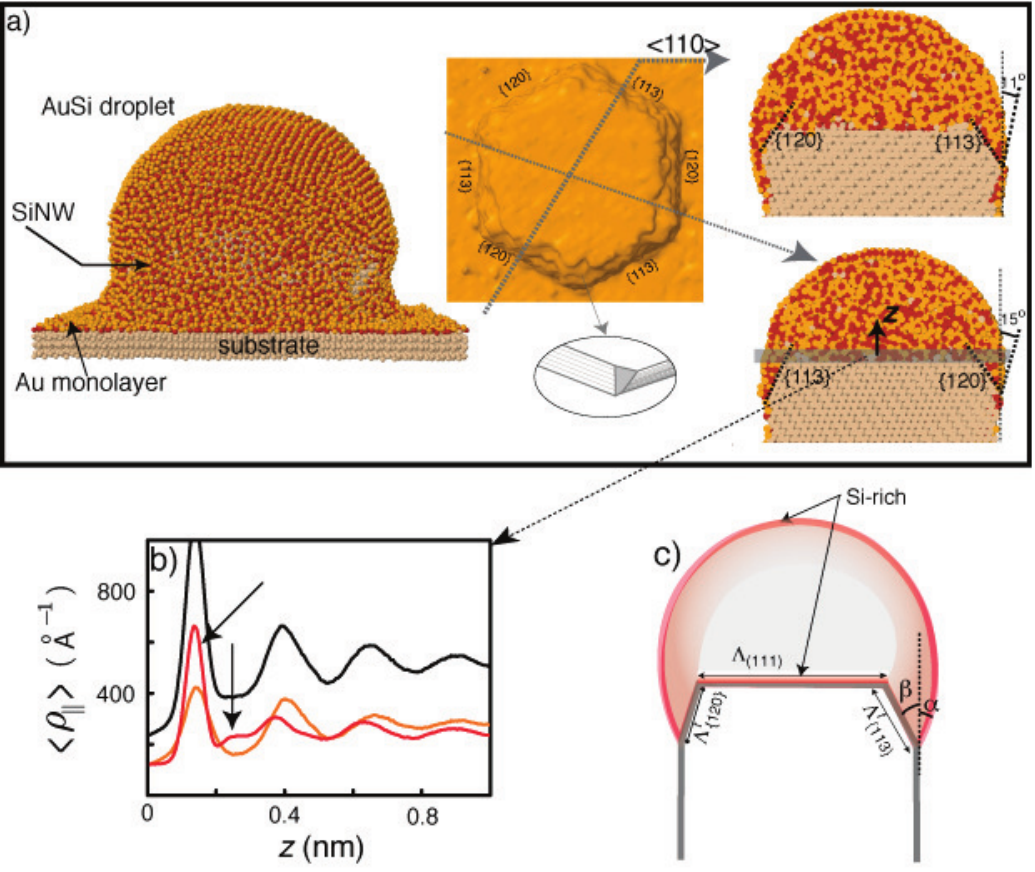}
\caption{\footnotesize {\bf Equilibrated droplet-nanowire-substrate system}: (left) Equilibrium atomic configuration of a $2R=10$\,nm diameter $\langle111\rangle$ silicon nanowire with $\{110\}$ sidewall facets. The wire is grown on a $(111)$ silicon substrate and is capped by an AuSi catalyst particle at the liquidus composition, $X_{Si}^\ast=0.46$. (middle, inset) The surface plot viewed along the $[111]$ direction showing the morphology of the solid SiNW that abuts the solid-liquid interface. The probe radius is fixed at $1.2\,{\rm \AA}$. Schematic illustration of one of the corners (arrow). The disordered region is shaded gray. (right) Atomic-plots of the midsections along the indicated directions showing the distribution of the $\{113\}$ and $\{120\}$ family of truncating facets flanking the interface. 
(b) The in-plane density profile  $\rho_\parallel$ averaged over each liquid bilayer abutting the $<$111$>$ interface [shaded strip in (a)] as a function of the distance $z$. The arrows indicate peaks consistent with Si segregation within the bilayer. (c) Schematic illustrating the interplay between the segregation profile and the interfacial energetics that maintains the local equilibrium at the contact line and the droplet shape, mainly angles $\alpha$ and $\beta$, main facet length $\Lambda_{(111)}$, and truncating facet lengths $\Lambda_{\{113\}}^t$ and $\Lambda_{\{120\}}^t$. 
\label{fig:figure3}
}
\end{figure*}

The substrate is explicitly included by mating an equilibrated droplet-SiNW system onto a dehydrogenated Si(111) surface (Methods). 
Figure~\ref{fig:figure3}a and Supplementary Figure S4 show the atomic configurations of the nanowire system with \{110\} and  \{112\} sidewall facets, respectively.  A stable Au submonolayer is evident on the nanowire sidewalls (right, Fig.~\ref{fig:figure3}a). The surface segregation is similar to that on isolated droplets (Fig.~\ref{fig:figure2}b), i.e. the substrate has a minor effect. There are differences in the structure of the Au submonolayer for the two classes of sidewall facets. We see Si segregation at the droplet-nanowire interface within liquid layers adjacent to the interface and parallel to $($111$)$ planes. Ripples in the in-plane density profiles within the droplet extracted as a function of distance from the interface $z$ extend to a thickness of $\approx5\,{\rm \AA}$ (Fig.~\ref{fig:figure3}b). The sharp Si and Au peaks adjacent to the interface at $z\approx1.5\,{\rm \AA}$ (arrow) indicate a Si-rich layer with some in-plane order. The remainder of the segregated layer is mostly liquid characterized by a smaller and broader Si-peak at $z\approx2.5\,{\rm \AA}$ (arrow), followed by a Au-rich region ($z\approx4\,{\rm \AA}$) which ultimately settles into the bulk composition.  The compositional distribution for the entire nanowire-droplet system is summarized schematically  in Fig.~\ref{fig:figure3}c.

Surface plot of the solid nanowire sans the the liquid atoms yields insight into the interface  morphology (middle panel, Fig.~\ref{fig:figure3}a). Clearly, it is not flat. The $\langle111\rangle$ facet is dominant yet it is truncated along the periphery next to the contact line. These truncated regions are 2-4 atomic diameters wide and their length varies non-uniformly along the contact line. Multiple midsections reveal a  stepped structure yet they are not smoothly curved; rather they correspond to two distinct orientations inclined at $\beta\approx32^\circ$ and $\beta\approx50^\circ$ to the $\langle111\rangle$ sidewall facet (vertical), which we identify to be the $\{113\}$ and \{120\} family of planes. Supplementary Figure S4 reveals a similar morphology for SiNWs with $\langle112\rangle$ sidewall facets with the exception that the truncating facets consist of \{113\} and \{111\} family of planes. The six facets alternate between these two inclinations along the contact line, as indicated in the surface plot and the midsections (Fig.~\ref{fig:figure3}a, right). The facets are also observed in the doubly-capped nanowire system shown in Fig.~\ref{fig:figure1}b. The six edge corners where truncating facets intersect are inherently rough as they have to absorb the differences in the facet inclinations, as illustrated schematically. The roughening can also be seen in the surface plot in that the step morphology kinks at the corner edges.  
The droplet shape is therefore asymmetric since the angles $\alpha$ and $\beta$ that follow from the Young's balance along the contact line are shaped by the facet energetics and geometry.
The asymmetry is also evident in the two midsections:  $\alpha\approx11^\circ$ at the $\{113\}$ truncating facet and changes to $\alpha\approx15^\circ$ at the $\{120\}$ truncating facet. 

The presence of nanometer-wide stable truncating facets is consistent with recent reports in several semiconducting nanowire systems, including the Au-Si system\cite{nw:OhChisholmRuhle:2010, nw:WenTersoffRoss:2011, nw:GamalskiHofmann:2011}. Limitations of the model system notwithstanding, such a complex 3D morphology composed of multiple families of truncating facets is similar to that recently reported in [0001] sapphire nanowires\cite{nw:OhChisholmRuhle:2010}, suggesting that this is perhaps a general feature during VLS growth. For the specific case of Si,  it is interesting that the $\{113\}$ and $\{111\}$  facets routinely decorate the \rm{sidewalls} during the complex sawtooth faceting in much larger diameter $\langle111\rangle$ SiNWs~\cite{nw:RossTersoffReuter:2005, nw:DavidGentile:2008}, suggesting that they perhaps nucleate first at the solid-liquid interface. 

\subsection{Nanowire growth kinetics}
\begin{figure*}
\centering
\includegraphics[width=1.75\columnwidth]{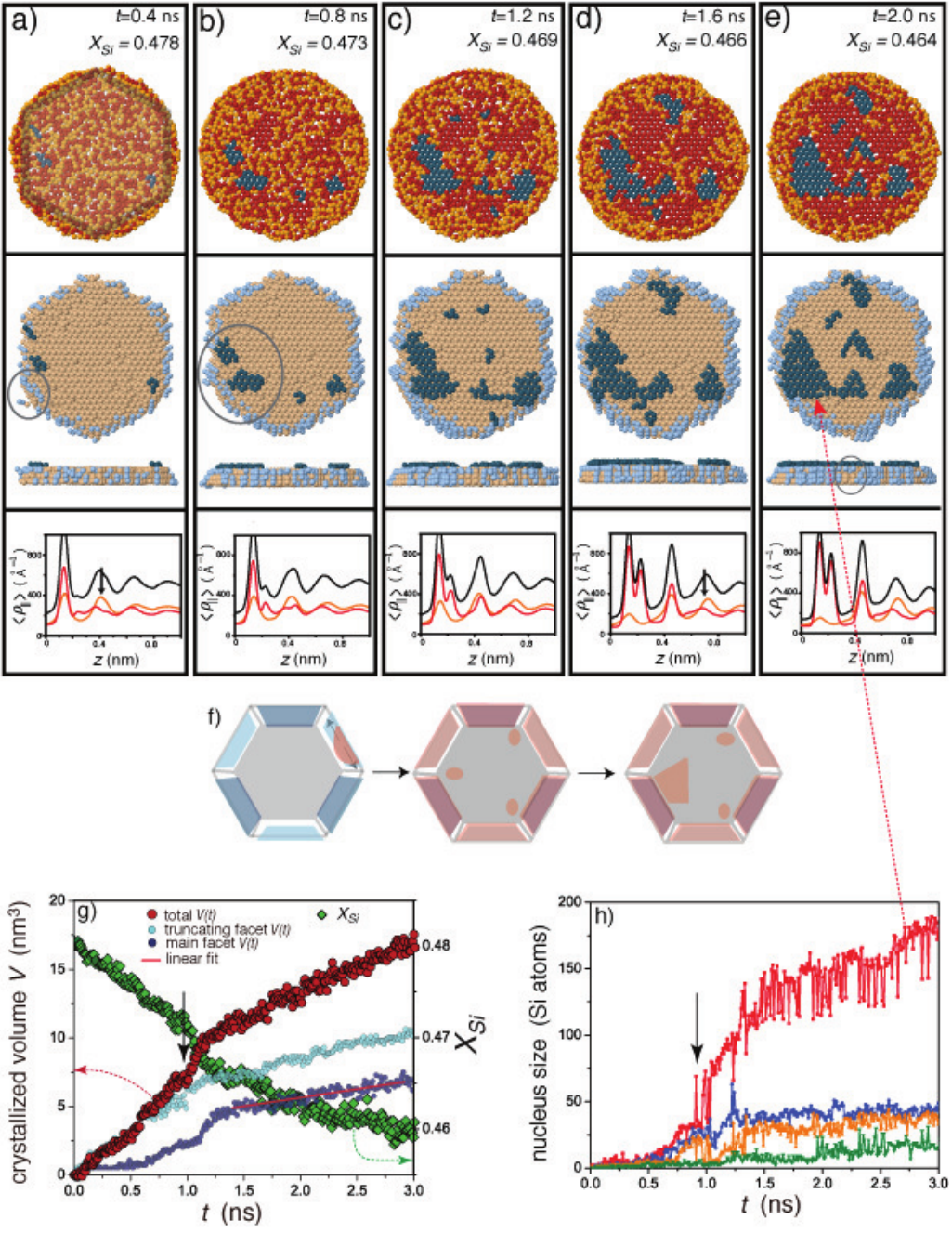}
\caption{\footnotesize {\bf Nucleation and growth kinetics at the nanowire-droplet interface}: (a-e) Atomic configurations in the vicinity of the interface subject to a supersaturated droplet with initial composition $X_{Si}=0.48$. (top row) Liquid bilayer abutting the droplet-SiNW interface, and (middle row) top and side views of the solid nanowire. The light and dark blue colors indicate crystallized Si atoms on the truncating and main facets, respectively. The instantaneous supersaturation is indicated on the top of each panel. The circled region in (a) highlights an instance of preferential growth at the center of the truncating facet while the circled region in (b) indicates precursor fluctuations that lead to the formation of a critical nucleus on the main facet. Details can be seen in Supplementary Movie 1.
The bottom of each panel shows in-plane density profile $\rho_\parallel(z)$. 
(f) Schematic illustration of the evolution of the interface before and during growth via step flow. See text for discussion.
(g) Temporal evolution of crystallized net Si volume (red circles) composed of crystallization on the edge and main facet (light and dark blue circles, respectively), as well as the droplet composition (solid green squares). The linear fit  (solid red line) used to extract the layer growth rate due to post-nucleation step flow on the main facet 
(h) Size of the nuclei observed on the main facet. The red curve is the evolution of the critical nucleus as indicated, while the other curves correspond to evolution of the smaller peripheral nuclei.
\label{fig:figure4}
}
\end{figure*}

The heterogeneous nature of interface facets has important ramifications for nanowire growth which we now study computationally. 
We simplify the catalysis by directly abstracting individual silicon atoms onto the droplet surface to a prescribed level of supersaturation, and employ classical molecular dynamics to study the non-equilibrium response of the entire droplet-SiNW-substrate system (Supplementary Methods). 
Figure~\ref{fig:figure4} shows the structural evolution at and around the interface as well as the in-plane density profile $\rho_{\parallel}(z)$ for a droplet with initial composition $X_{Si}=0.48$ ($\Delta_T\approx36$\,K). 
The initial growth stage for $t<0.4$\,ns is marked by layer-by-layer crystallization on the truncating facets. 
The {\it circumferential} growth is likely aided by their small widths and the proximity to the Si-rich contact line (circled, Fig.~\ref{fig:figure4}a).  We see 1D lengthening of kinks associated with the stepped structure of these facets.  Interestingly, the growth occurs preferentially away from the corner edges implying that they favor dissolution. For details, see Supplementary Movie 1. The extent of truncation changes due to their growth, consistent with the oscillatory growth wherein the facet widths decrease between bilayer additions on the main facet\cite{nw:OhChisholmRuhle:2010, nw:WenTersoffRoss:2011, nw:GamalskiHofmann:2011}.  We also see smaller clusters of 2-3 Si atoms that continually form and dissolve on the main facet yet the density profile normal to the interface remains unchanged (bottom row), i.e. the crystallization is limited to the truncating facets. 

The $t=0.8$\,ns configuration plotted in Fig.~\ref{fig:figure4}b  shows that, barring the corner edges, the growth has advanced by almost a full layer on the truncating facets. The droplet is still supersaturated
and it now drives the nucleation on the main facet.  It is worth pointing out that structurally the growth on the truncating facets and on the main facet is essentially independent, since a 2D island reaching the edge of one facet has no nearest neighbors with an island on the adjacent facet. They require the influence of the much weaker next nearest-neighbor interactions to extend across. We observe early stages of classical 2D layer-by-layer growth wherein several nuclei form and attempt to overcome the nucleation barrier. Observe that the nuclei form slightly away from the center of the main facet and their location is correlated with the corner edges. The disorder together with the enhanced Si-segregation at the contact line likely aids the nucleation on the main facet.
The diffusion along the surface is indeed slower yet the bulk liquid can easily provide a pathway.  The precursor cluster that eventually becomes supercritical is circled in Fig.~\ref{fig:figure4}b. Figures~\ref{fig:figure4}c-\ref{fig:figure4}e reveal the key features of the subsequent bilayer growth on the main facet, i.e. it remains layer-by-layer and dominated by a single growing nucleus. 

The density profiles evolve accordingly. The intensity of the two silicon peaks increases at the expense of the Au peak.  A sharp peak in the adjacent liquid layer indicates build-up of Si while the third layer becomes Au-rich (arrow in Fig.~\ref{fig:figure4}d). Evidently, Au diffuses rapidly away from the interface. At $t=2$\,ns, the first layer has not completely formed yet the step flow has slowed down due to the low supersaturation left in the droplet, $X_{Si}=0.463$. Note that there is negligible growth at the edge corners (circled, Fig.~\ref{fig:figure4}e).
We see Au and liquid Si within the mostly crystallized Si layer although the liquid Si atoms exhibit in-plane order. Eventually, some of the Au atoms diffuse away into the liquid droplet and aid further crystallization (not shown). 

We quantify these observations by monitoring the crystallized volume at the interface $V(t)$ and the droplet composition $X_{Si}(t)$ (Fig.~\ref{fig:figure4}g). 
Crystallization kinetics on the truncating and main facets are plotted separately. Initially for $t<1$\,ns the increase in $V(t)$ is entirely due to crystallization on the truncating facets. A linear fit yields a growth rate of $v\approx10.5$\,cm/s. The growth is instantaneous indicative of a small nucleation barrier. While the barrier can be artificially reduced due to the larger supersaturations in the computations, note that the Si-rich contact line is an easy source for Si adatoms. Also, as in the experiments, the facet widths rarely exceed a few nanometers such that the critical nuclei sizes are likely larger than the facet widths\cite{sold:RohrerMullins:2001}. The circumferential 1D growth is also aided by the stepped structure of these facets and we quantify this effect by extracting Si adatom enthalpies $H_{ad}$ on the $\{111\}$ and the $\{113\}$ family planes (Methods). Not surprisingly, the close-packed $\{111\}$ facets have a significantly higher enthalpy than the stepped $\{113\}$ facets, $\Delta H_{ad}\approx2.0$\,eV/adatom. The adatom diffusivity on the $\{113\}$ facets is the suppressed since the atom makes fewer diffusion hops before another atom crystallizes in close proximity. The adatoms can readily form clusters and therefore lower the nucleation barrier, which in turn can induce rapid 1D growth. The role of such structural effects in introducing kinetic anisotropies during faceted crystal growth is well-established\cite{cg:ZepedaRuizGilmer:2006}. 

At $t\approx1.0$\,ns, the growth rate on the truncating facets decreases, either due to site-saturation or the onset of the oscillatory growth mode.
The growth on the main facet, non-existent so far, dramatically increases. This is indicative of the growth of a supercritical nuclei and is again consistent with our qualitative observations (Fig.~\ref{fig:figure4}c-\ref{fig:figure4}e). In order to ascertain if the growth is layer-by-layer or kinetically rough, we monitor the size evolution of the visibly identifiable nuclei on the main facet (Fig.~\ref{fig:figure4}h). We see a dramatic increase in the size of one of the nuclei while the other nuclei fluctuate about a relatively smaller size ($<35$ atoms). At longer times $t\approx10$\,ns, one of these smaller nuclei shrinks while the other two persist without any appreciable increase in size. Clearly, one 2D island dominates the layer growth and the encompassing steps are the main contributions to surface roughening. Moreover, the step orientations reflect the symmetry of the growth direction while that is not the case for the fluctuating subcritical islands. The size evolution shows that even though the driving force is larger than that in experiments, kinetic roughening or the increase in step density resulting from growth on the main facet is non-existent\cite{sold:JacksonGilmer:1976}. The persistence of the subcritical nuclei can signify that the supersaturation is on the borderline between single nucleus and polynuclear growth; then at lower supersaturations we expect the growth of the nanowires to be universally by the single nucleus mechanism. Following nucleation, the growth is steady-state as the volume increases and composition decreases. A linear fit in this regime yields the overall growth velocity, $v\approx6.3$\,cm/s. The growth rate also includes contributions from the truncating facets. Subtracting the former from the overall growth rate yields the step velocity on the main facet, $v_s\approx3$\,cm/s. Figure~\ref{fig:figure4}f is a schematic summary of the morphological features implicated in the overall growth.

\subsection{Effect of supersaturation}
\begin{figure*}
\begin{center}
\includegraphics[width=1.5\columnwidth]{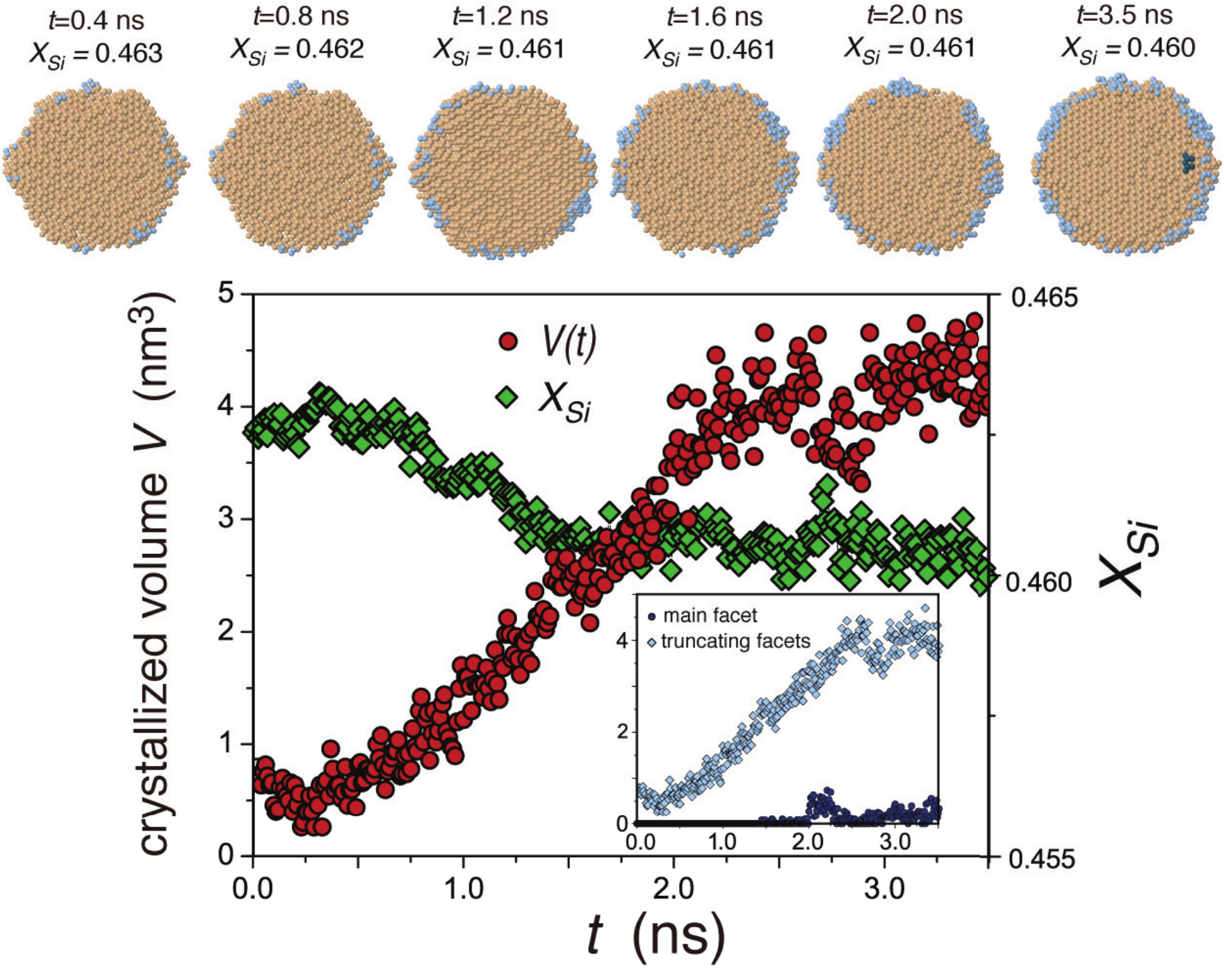}
\end{center}
\caption{\footnotesize {\bf Growth kinetics at small supersaturations}: (top row) Atomic configurations of the solid nanowire growth front subject to an initial droplet supersaturation of $X_{Si}=0.463$. The elapsed time and the current supersaturation are also indicated. For details, see Supplementary Movie 2.
(bottom) The temporal evolution of the crystallized volume $V(t)$ and the composition within the droplet, $X_{Si}(t)$. (inset) Evolution of the crystallized volume on the main and truncating facets. 
\label{fig:figure5}
}
\end{figure*}

In order to make contact with experimental scales,  we quantify the nucleation barrier on the main facet by performing computations with decreasing initial supersaturation. Representative results for $X_{Si}=0.463$ ($\Delta_T\approx5.4$\,K) are shown in Fig.~\ref{fig:figure5}. The truncating facets again grow rapidly and circumferentially (Fig.~\ref{fig:figure5}a). The growth is instantaneous yet slower, and the growth on the main facet is non-existent (Fig.~\ref{fig:figure5}b). The crystallization is again preferentially at the centers of the truncating facets, presumably due to a 1D interplay between crystallization at the facet center and melting at the rough corner edges (Supplementary Movie 2). At $t\approx3.0$\,ns, we see transient crystallization of subcritical nuclei on the main facet yet there is no supersaturation left for their growth. Then, $X_{Si}=0.463$ represents a critical initial composition below which there is no nucleation on the main facet, i.e. nanowire growth can only occur at larger supersaturations.

Increasing the initial composition in the range $X_{Si}\ge0.485$ ($\Delta_T\ge45$K) decreases the dead-zone that precedes the nucleation on the main facet. Multiple stable islands contribute to the growth of a single layer such that the step density increases and there is a gradual transition to a kinetically roughened regime wherein the second bilayer begins to crystallize before the complete growth of the first bilayer. This is evident in configurations for $X_{Si}=0.485$ shown in Supplementary Figure S5.  Several Au atoms remain confined within the first layer. They likely retard the step flow and that triggers the onset of a 3D growth mode. Midsections reveal additional Si diffusion down the sidewalls that further aids the absorption of the supersaturation  (inset).

\begin{figure*}
\centering
\includegraphics[width=2\columnwidth]{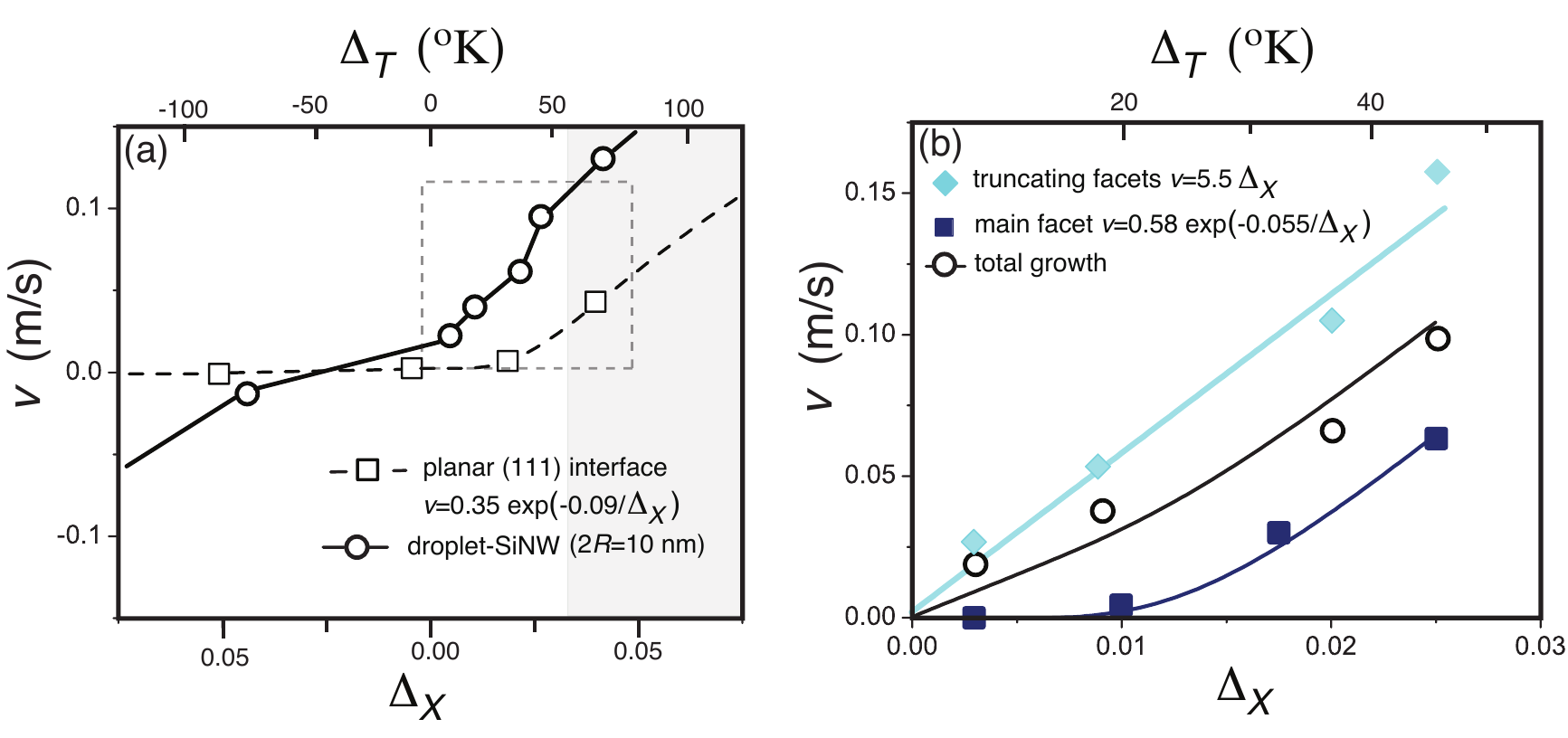}
\caption{\label{fig:figure6} \footnotesize {\bf Supersaturation dependence of nanowire growth rate}: (a) The extracted linear growth velocity $v[=1/(\pi R^2)\,dV/dt]$ of the nanowire as a function of the initial supersaturation, $\Delta_X=X_{Si}-X_{Si}^\ast$. The equivalent undercooling based on the phase diagram in Fig.~\ref{fig:figure1}a is also indicated. 
For comparison, growth velocity of a planar Si(111)-AuSi  solid-liquid interface is also plotted (dotted line) as a function of initial supersaturation $\Delta_X(R=\infty)$ within the AuSi phase. The gray shaded areas represent kinetically roughened growth regime induced by large supersaturations. (b) Expanded view of the small supersaturation growth data corresponding to the boxed region in (a). The contributions of the two classes of facets and the relevant curve fits are shown as separate plots. 
}

\end{figure*}

Figure~\ref{fig:figure6}a shows the driving force dependence of the overall growth velocities for varying supersaturations. The velocities are linear fits to the initial crystallization rates. 
In the small driving force limit $X_{Si}\le0.48$ ($\Delta_T<36$\,K), the velocity increases rapidly. Results of similar computations on a planar Si(111)-AuSi interface are also plotted for comparison (see Methods).  Crystallization at the planar interface entails homogeneous nucleation and is always slower at comparable driving forces, suggesting a higher nucleation barrier. 
Separate plots for the contributions of the truncating and main facets are plotted in Fig.~\ref{fig:figure6}b. The velocity of the truncating facets increases linearly with supersaturation and indicates that its growth is not nucleation-controlled, at least for undercoolings as low as $\Delta_T=5.4$\,K. The average kinetic coefficient is $\approx3.0$mm/(s\,K) and is of the order of kinetic coefficients in binary alloys\cite{sold:HoytAstaKarma:2002}.

The crystallization rate on the main facet increases non-linearly with the supersaturation signifying classical nucleation-controlled growth. It is possible that at much smaller driving forces the growth is linear as the finite facet size requires equilibrium between melting and nucleation\cite{nw:HaxhimaliAstaHoyt:2009}, although it is not clear if this applies to a fully faceted interface.  
Accordingly, we fit the variation to a functional form $v=v_0\exp(-\Delta G/k_BT)$, where the exponential factor is the probability that a spontaneous fluctuation will result in a critical nucleus, and $v_0$ is the 2D growth rate of the supercritical nucleus that is almost independent of the supersaturation. The nucleation barrier varies as $\Delta G \propto 1/\Delta_X$ 
and the exponential fit to the extracted velocities yields a value $\Delta G=4$\,meV/$\Delta_X$ at $T=873$\,K.  A similar fit to the planar interface data yields a barrier that is $63\%$ larger and is consistent with estimates based on nucleation of a 2D island (Supplementary Discussion 1). At larger driving forces characterized by kinetically rough 3D growth, the velocity increases almost linearly with supersaturation.  The plot also shows results of computations on undersaturated droplets ($X_{Si}<X_{Si}^\ast$). At small driving forces, the nucleation and growth dynamics associated with the melting kinetics is completely reversed; the truncating facets melt first as the energy of formation of a vacancy on these facets is lower, and eventually the main facet melts preferentially from the corner edges. 
The layer-by-layer  melting velocities are smaller compared to those during growth due to the
 different morphological features implicated in the two processes, which by itself is important as the growth is a steady-state between growth and melting kinetics\cite{sold:RohrerMullins:2001, nw:HaxhimaliAstaHoyt:2009}.

\subsection{Mechanistic insight during growth}
\begin{figure*}
\centering
\includegraphics[width=1.5\columnwidth]{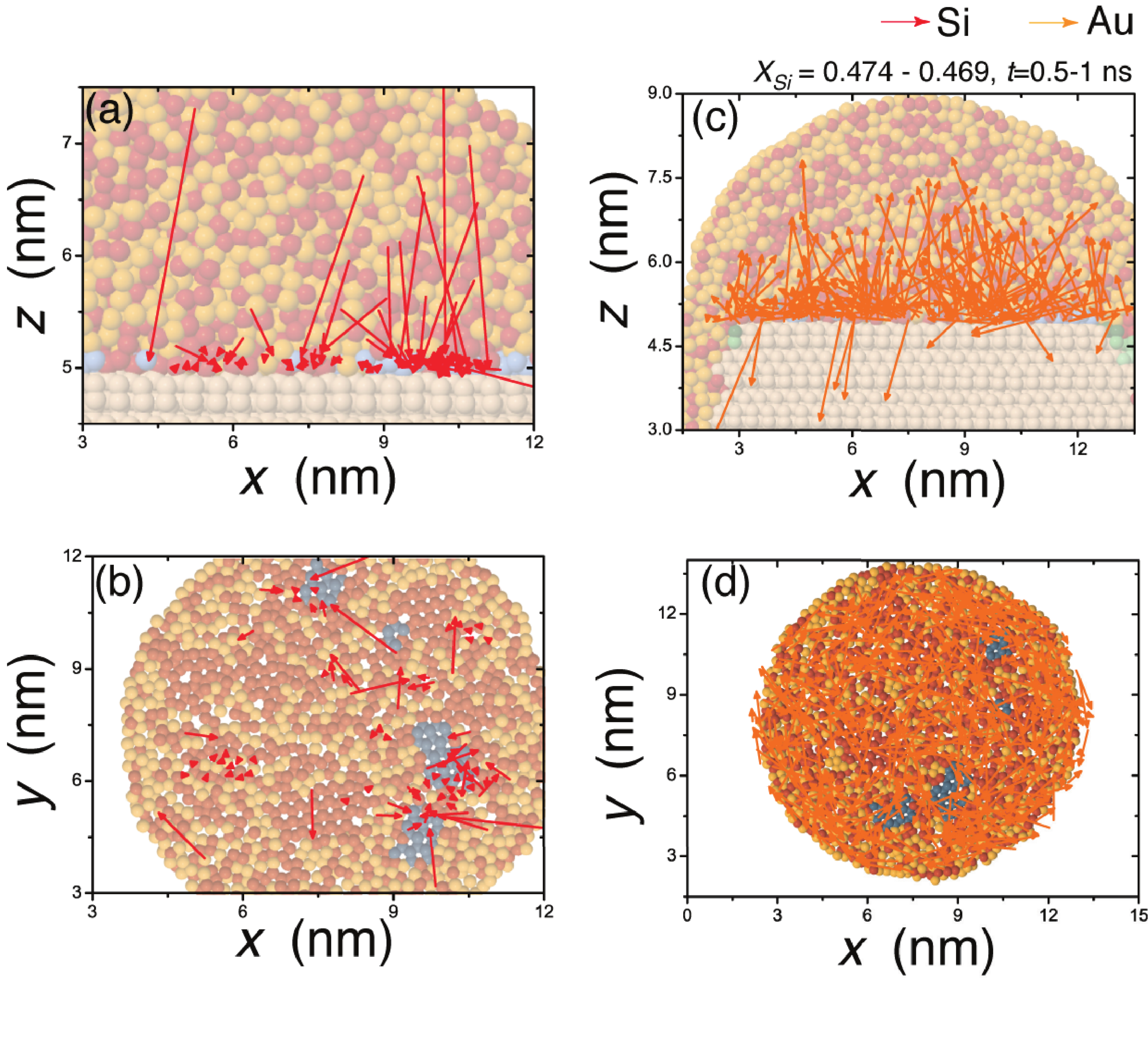}
\caption{\label{fig:figure7} \footnotesize {\bf Diffusion pathway during nanowire growth}: (a-d) Cumulative displacement maps for select (a-b) Si and (c-d) Au atoms over a $0.5$\,ns time interval following nucleation, for initial composition $X_{Si}=0.48$. Both side (a, c) and top (b, d) are shown. The Si displacements correspond to crystallized atoms within the time interval. The Au displacements are for atoms initially at the solid-liquid interface. 
Within each map, the arrow lengths are proportional to the net displacement.  Atomic configurations of midsections (side views) and bilayers (top views) are superposed within the maps to facilitate visual location of the crystallizing Si atoms (shaded blue).
}
\end{figure*}
Maps of atomic displacements over prescribed time intervals corroborate the observed trends. Figure~\ref{fig:figure7}a-d shows a $t=0.5-1.0$\,ns map for select Si and Au atoms within the $X_{Si}=0.48$ computation. The displacements correspond to newly crystallized Si atoms (Si-map) and Au atoms initially at the interface (Au-map). The 2D step flow results in linear decrease of the droplet composition from $X_{Si}=0.474$ to $X_{Si}=0.469$ (Fig.~\ref{fig:figure4}g).
The Si crystallization is mostly from adjacent bilayers and driven by displacement chains. Surprisingly, we also see long-range fast diffusion from as far as the droplet surface (Fig.~\ref{fig:figure7}a). The  flux is mostly through the amorphous bulk of the droplet and we see negligible surface diffusion (Fig.~\ref{fig:figure7}b). The step flow is aided by Au diffusion into the droplet (Figs.~\ref{fig:figure7}c) and also along the interface directed away from the moving steps (Fig.~\ref{fig:figure7}d). We see evidence for both short-range atomic hops into adjacent liquid bilayer and long-range diffusion towards the droplet surface. The diffusion is markedly along the bulk and also towards the nanowire sidewalls; the latter represents a direct link between nanowire growth and Au decoration on the sidewalls.

\section*{Discussion}
The atomic-scale structure and dynamics during VLS growth of SiNWs is clearly sensitive to segregation at the surface and the droplet-nanowire interface.
The robust surface layer 
is expected to be stable under reactor conditions where the droplet is exposed to the precursor gas atmosphere. The Au submonolayer together with the inhibited surface diffusion renders the droplet surface in immediate contact with the vapor sufficiently Au-rich to drive the precursor breakdown over large pressure and temperature ranges. 
The heterogeneously faceted interface morphology impacts all aspects of the growth process\cite{nw:WenTersoffRoss:2011} and requires reconsideration of current growth models. The truncating facets induce variations in the Si segregation and droplet morphology along the contact line. The relative disordered corner edges impact interface nucleation and growth. They likely serve as efficient sources of steps for rapid crystallization on the truncating facets, analogous to the screw dislocation-based spiral crystal growth\cite{tsf:BurtonCabreraFrank:1951}. At much lower supersaturations, the dynamics may involve a competition between nucleation at the facet centers and melting at the edge corners. Nevertheless, the barrier is clearly much smaller than that on the main facet. The latter is clearly the rate limiting event. The growth of the truncating facets also decreases the extent of these ordered regions, and they can facilitate mass transfer from the Si-rich contact line onto the main facet and thereby lower the barrier compared to an equivalently supersaturated planar Si-AuSi interface; the computations show that this is indeed the case. The extracted nucleation barrier $\Delta G$ allows us to establish nucleation- and catalysis-controlled regimes. Under typical growth conditions, the atomic incorporation rate is of the order of a single Si atom per millisecond (Supplementary Discussion 2). Equating this incorporation rate to the nucleation controlled growth rate, the extracted barrier $\Delta G=4$\,meV$/\Delta_X$ yields a supersaturation of $\Delta_X^c\approx0.002$ for isothermal nanowire growth at $T=873$\,K, or a critical undercooling of $\Delta_T^c\approx3.6$\,K. The growth is then nucleation-controlled for $\Delta_T<\Delta_T^c$ and catalysis-controlled otherwise. Solidification from melts typically occurs at undercooling of less than a Kelvin\cite{book:Woodruff:1973}, and if that is the case during VLS growth of nanowires, it must nucleation-controlled. We should emphasize that since the barrier is expected to be nanowire size dependent, the interplay between nucleation and catalysis can be quantitatively different at larger sizes and is a focus of future studies. 

An interesting coda is the transition to kinetically rough growth mode at larger supersaturations is intriguing as the step flow is increasingly limited by Au-diffusion away from the step-edges. The inter-island Au atoms are stable over the MD time scales ($<10$\,ns), and although we cannot rule out their slow diffusion out of the nanowire eventually, there is growing evidence of catalyst particle trapping within the as-grown nanowires\cite{nw:MoutanabbirSeidman:2013}. Detailed understanding of this regime opens up the possibility of doping nanowires to non-equilibrium compositions {\it during} their synthesis\cite{nw:SchwalbachVoorhees:2009}, a scenario worthy of detailed explorations by itself. 

\section{Methods}
{\bf Model system}: We use classical inter-atomic potentials to model the eutectic AuSi system. Hydrogenation effects are ignored due to hydrogen desorption and Au-surface passivation at these elevated temperatures. The potentials can be a limitation as they represent approximations to the electronic degrees of freedom. Although the AEAM model system is fit to several experimental and first-principles data and is based on established frameworks for both pure Au and Si,
its ability to make quantitative predictions, notably interface properties, may be limited. For example, the pure Au potential underestimates the liquid free energies\cite{intpot:WebbGrest:1986}  while the pure Si potential is limited in describing surface reconstructions on crystal facets\cite{intpot:LiBroughton:1988}. Nevertheless, validation studies presented here show that the predictions of the model potential are in agreement with several alloy properties. The AEAM framework reproduces the primary feature of the binary phase diagram, i.e. a low melting eutectic.  The noteworthy deviations are an underestimation of the eutectic temperature $T_E$ ($590$\,K compared to $636$\,K) and overestimation of the eutectic composition $X_{Si}^{E}$ ($31\%$ compared to $19\%$). We have also performed contact angle studies of eutectic and hypereutectic AuSi droplets  on Si(111) surface using semi-grandcanonical Monte-Carlo (SGMC) simulations (Supplementary Methods). The final configuration of a $2R=10$\,nm size droplet on a Si(111) surface at the growth temperature is shown in Supplementary Figure S6. The contact angle $\theta=143^\circ$ is close to the experimental value\cite{tsf:ResselHomma:2003}, $\theta=136^\circ$.  

{\bf Equilibrium liquidus concentration for $2R=10\,$nm SiNW}: The initial structure consists of a pristine $2R=10\,$nm $\langle111\rangle$ SiNW with $\{110\}$ or $\{112\}$ family of lateral facets capped by two AuSi particles. The Si(111)-AuSi interfaces are initially flat. The SiNW length is $\approx8$\,nm, large enough such that the interfaces do not interact. The atomic structure of the particles is generated randomly at a density corresponding to liquid AuSi and the initial composition is hypoeutectic, $X_{Si}=0.25$. The entire system consists of $\sim64000$ atoms and is relaxed using SGMC computations.
A local order parameter is employed to demarcate the liquid and solid Si atoms\cite{sold:ButaAstaHoyt:2008}. The order parameter is an extension of the pure Si case based on the local 3D structure around each Si atom. Additional validation of the surface segregation is done using charge-gradient corrections to the Au-Au interactions that correctly reproduces the liquid Au surface tension (Supplementary Methods).

{\bf Equilibrium studies on isolated AuSi particles and thin films}:
Density profiles of the equilibrated droplets 
are averaged over bilayer thick spherical shells around the center of mass of the droplet, and the ensemble average is over 1000 different atomic configurations. In order to quantify the effect of droplet surface curvature, the density profiles are also extracted for thin films. We employ a thin film of thickness $H=10\,$nm, free at both surfaces and periodic in the film plane. SGMC simulations are used to equilibrate the initially liquid film. Supplementary Figure S1 shows the equilibrated configuration for a 27000-atom simulation and the corresponding normalized density profiles as a function of a distance from the surface $h-H$ extracted at the bulk eutectic composition and temperature, $X_{Si}=0.31$ and $T_E=590$\,K. 
Layering of both Au and Si is clearly visible and a Au monolayer decorates the Si-rich surface. 
Supplementary Figure S1 also shows the density profiles for thin films at the growth temperature selected for this study ($T=873$\,K) and for size-corrected liquidus point, ($T=873$\,K, $X_{Si}=0.46$).  Comparison with the profile at the eutectic point clearly shows a decrease in the thickness of the solid surface layer with increasing temperature and Si supersaturation.

{\bf Au and Si diffusivities within the droplet}: MD simulations are performed on the equilibrated droplets to extract the ensemble-averaged MSD of both Au and Si (Supplementary Methods). A net simulation time of a few tens of picoseconds is sufficient to extract statistically meaningful data. For each atomic species at an initial distance $r$ from the mass center of the particle,  the MSD increases linearly with time and the slope yields the diffusivity$D(r)$.  For atoms in the bulk and on the surface, the slope is different. The MSDs for Au and Si atoms initially at the surface of the $2R=10$\,nm Au$_{54}$Si$_{46}$ droplet are plotted in Supplementary Figure S3. The computations on bulk eutectic AuSi alloy yield a diffusivity of $D_{Au}=1.54\times10^{-9}$m$^2$/s at $873$\,K; the value is in excellent agreement with past experiments on Au diffusivity in eutectic AuSi alloys\cite{diff:BrusonGerl:1982}, $D_{Au}\approx1.7\times10^{-9}$m$^2$/s. 

{\bf Equilibrium state of the droplet-nanowire-substrate system}:
The initial configuration is generated by taking one half of the double droplet configuration and placing it on a ($16.0\,{\rm nm}\times15.4\,{\rm nm}$) Si(111) substrate. The dimensions of the substrate are the same as in the contact angle studies. The bottom bilayer is fixed. The exposed area is decorated by a Au monolayer with a structure similar to the classical $\sqrt{3}\times\sqrt{3}\,\uppercase{R}\,30^\circ$ monolayer structure\cite{tsf:NagaoHenzler:1998}, and the entire system is equilibrated using canonical Monte-Carlo simulations followed by MD simulations. 

{\bf Nanowire growth kinetics}:
The catalytic incorporation of individual Si atoms takes place over much larger time-scales, and we expect the interface growth to be orders of magnitude faster.
Therefore, there is no additional Si flux imposed on the droplet during the course of the MD computations and we investigate the effect of an initially prescribed driving force that evolves in a manner consistent with nucleation-limited growth. Although the supersaturation is arbitrary and likely larger compared to the critical supersaturation in experiments, a systematic variation of the driving force allows us to extract the key features associated with the near-equilibrium behavior of the nanowire during its growth.
The supersaturation for growth is generated by attempting to place Si atoms on the surface within an MD simulation and then accepting the move if an Au atom is within a nearest-neighbor distance. Alternate procedure wherein the droplet is supersaturated by directly introducing atoms within an isolated droplet have negligible effect on the ensuing dynamics; the segregation at the surface and interface is fairly robust. The MD simulations employed are same as those for diffusivity calculations. 

\subsection*{Si(111)-AuSu interface kinetics and step energetics}
The simulation cell contains of a crystalline Si(111) substrate (8640 atoms) abutting a AuSi alloy (34560 atoms). The dimensions of simulation cell is $14.14 a_0\times14.69 a_0\times 25.97a_0$, where $a_0$ is the lattice parameter at $873$\,K. The initial AuSi alloy with the desired concentration is randomly structured and MD simulations are performed at $873$\,K. 
Periodic boundary conditions are applied along the in-plane directions with free surfaces normal to the interface. 
The Si(111) substrate is initially fixed and the AuSi alloy is relaxed for $0.1$\,ns. The planar nucleation/melting simulations are carried by fixing the bottom bilayer of the substrate.
The crystalline and liquid Si atoms are identified using the local order parameter described earlier.
The simulation cell for extracting the step enthalpies is similar; it $19.80 a_0\times19.60a_0\times10.39 a_0$ in size and consists of a Si(111) slab (16128 atoms) and a AuSi alloy slab (16128 atoms). The composition of the initial AuSi alloy is $X_{Si}=0.46$ and it is relaxed using isobaric, isothermal MD simulations (Supplementary Methods). The simulation cell is fully periodic so that the cell consists of Si(111)-AuSi interfaces. The dimension normal to the free surface adjusts to maintain zero pressure while two in-plane dimensions $(1\bar{1}0)$ and $(11\bar{2})$ are fixed. Following relaxation ($0.3$\,ns), the total potential energy $U_1$ is averaged over $0.1$\,ns. Single bilayer steps are created on both surfaces of Si slab while preserving the atoms in the Si(111) crystal and liquid AuSi. The steps are oriented along the index $(112)$ direction. The total potential potential energy is recorded as $U_2$ and the difference $\sigma=(U_2-U_1)/L=0.16$\,eV/nm, where $L$ is the total step length.

{\bf Si adatom energetics on truncating and main facets}:
Canonical MD simulations are performed at temperature $T=1$\,K.
Each computational cell consists of an unreconstructed crystal Si slab, free at both both surfaces and periodic in the slab plane. The dimensions of simulation cells are [$7.07 \times 9.38 a_0 \times 6.63 a_0$] and [$7.07 a_0 \times 7.35 a_0 \times 5.20 a_0$] for $(113)$ and $(111)$ surfaces and contains 3520 and 2160 atoms, respectively. The slabs were relaxed for $10$\,ps. A single Si atom is placed within the interaction range on each of the surfaces and relaxed for another 10 ps. All possible positions with maximum nearest neighbors were simulated to obtain the lowest potential energy. The maximum potential energy difference before and after addition of the Si adatom is recorded as the Si adatom enthalpy $H_{ad}$. The extracted enthalpies are $-5.4$\,eV/atom and $-3.5$\,eV/atom for the $(113)$ and $(111)$ surfaces, respectively. 
Note that the stepped structure of unreconstructed $(113)$ surfaces can be interpreted as $(111)$ steps separated by single atomic rows of $(001)$ terraces. The $\{120\}$ facets are also stepped and we therefore expect similar trends.

\noindent
{\bf Acknowledgements}: We thank A. Dongare and L. Zheiglei for providing us with an early version of the AEAM interatomic potential, and Alain Karma and Albert Davydov for helpful discussions. The computations were performed on {\it st}AMP supercomputing resources at Northeastern University. The study is supported by National Science Foundation DMR CMMT Program (1106214). Part of this work was performed under the auspices of the U.S. Department of Energy by Lawrence Livermore National Laboratory under Contract DE-AC52-07NA27344 (LAZ and GHG).




\clearpage

\begin{widetext}
\section*{\Large Supplementary Information}

\section*{Supplementary Figures}

\begin{figure}[h!tb]
\begin{center}
\includegraphics[width=0.6\columnwidth]{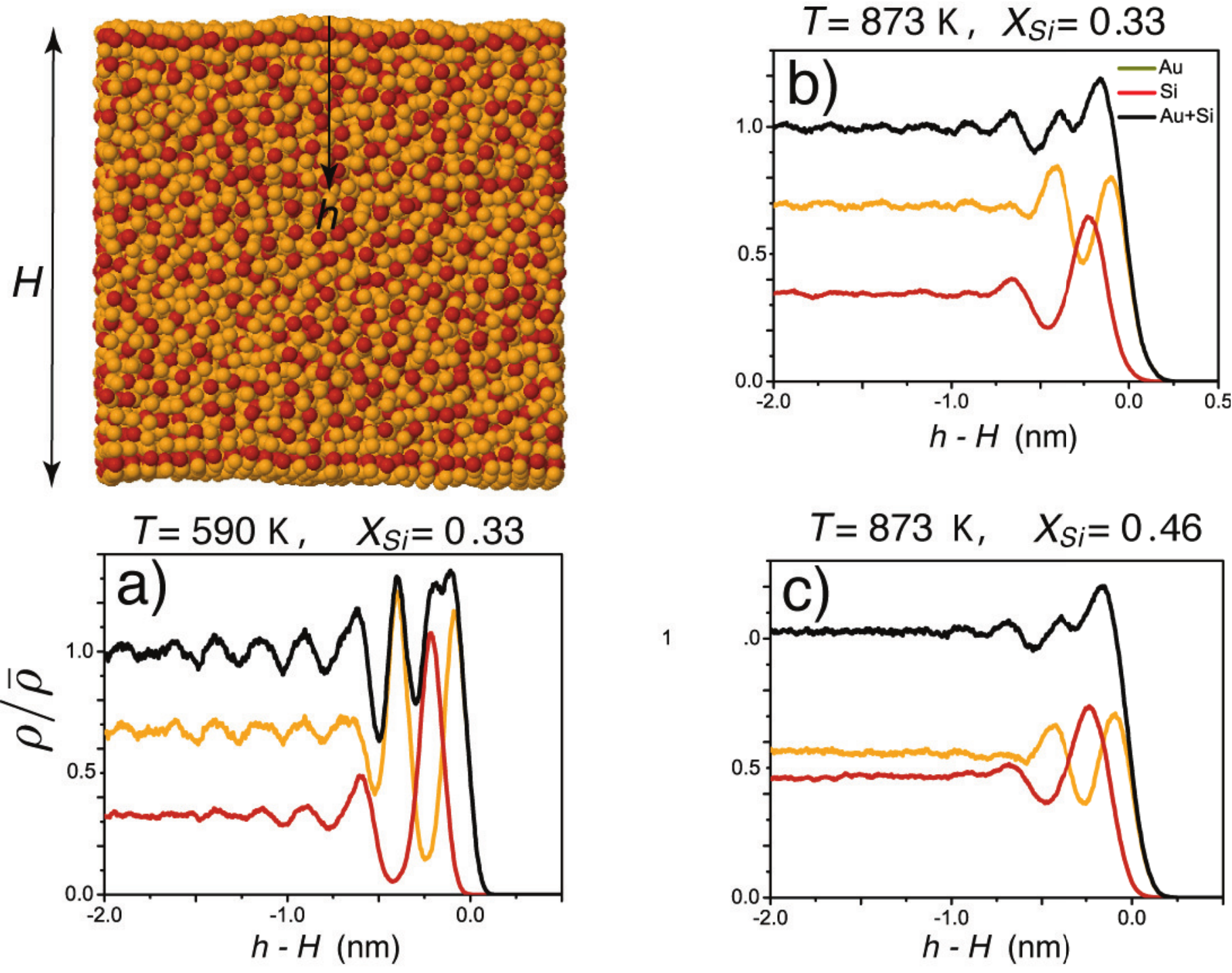}
\end{center}
\caption{(top left) Atomic configuration and the corresponding normalized density profiles as a function of distance from the surface ($h-H$) in an AuSi thin film at the eutectic point for the model system. (a) The normalized density profiles at (a) eutectic point $X_{Si}^\ast(590\,{\rm K})=0.31$, (b) elevated temperature $T=873$\,K, and (c) elevated temperature and supersaturation, $X_{Si}(873\,{\rm K})=0.46$.
\label{fig:Supp2}
}
\end{figure}

\begin{figure}[h!tb]
\begin{center}
\includegraphics[width=\columnwidth]{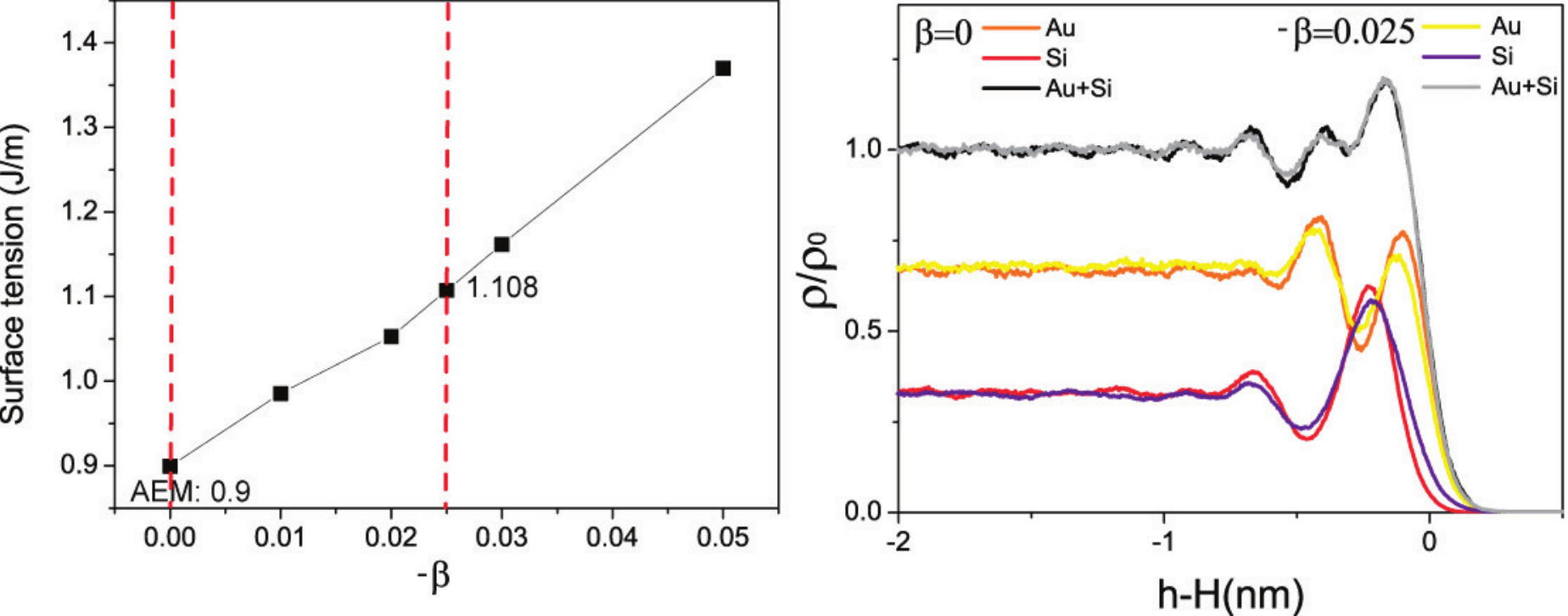}
\end{center}
\caption{\label{fig:Supp5}  a) Surface tension $\gamma$ versus $-\beta$ for AEAM Au at 1400\,K. $\beta=-0.025$ is a fit to the experimental value. (b) Segregation profiles for the AEAM Au/Si alloy thin film,  before and after charge gradient correction.}
\end{figure}

\begin{figure}[h!tb]
\begin{center}
\includegraphics[width=0.55\columnwidth]{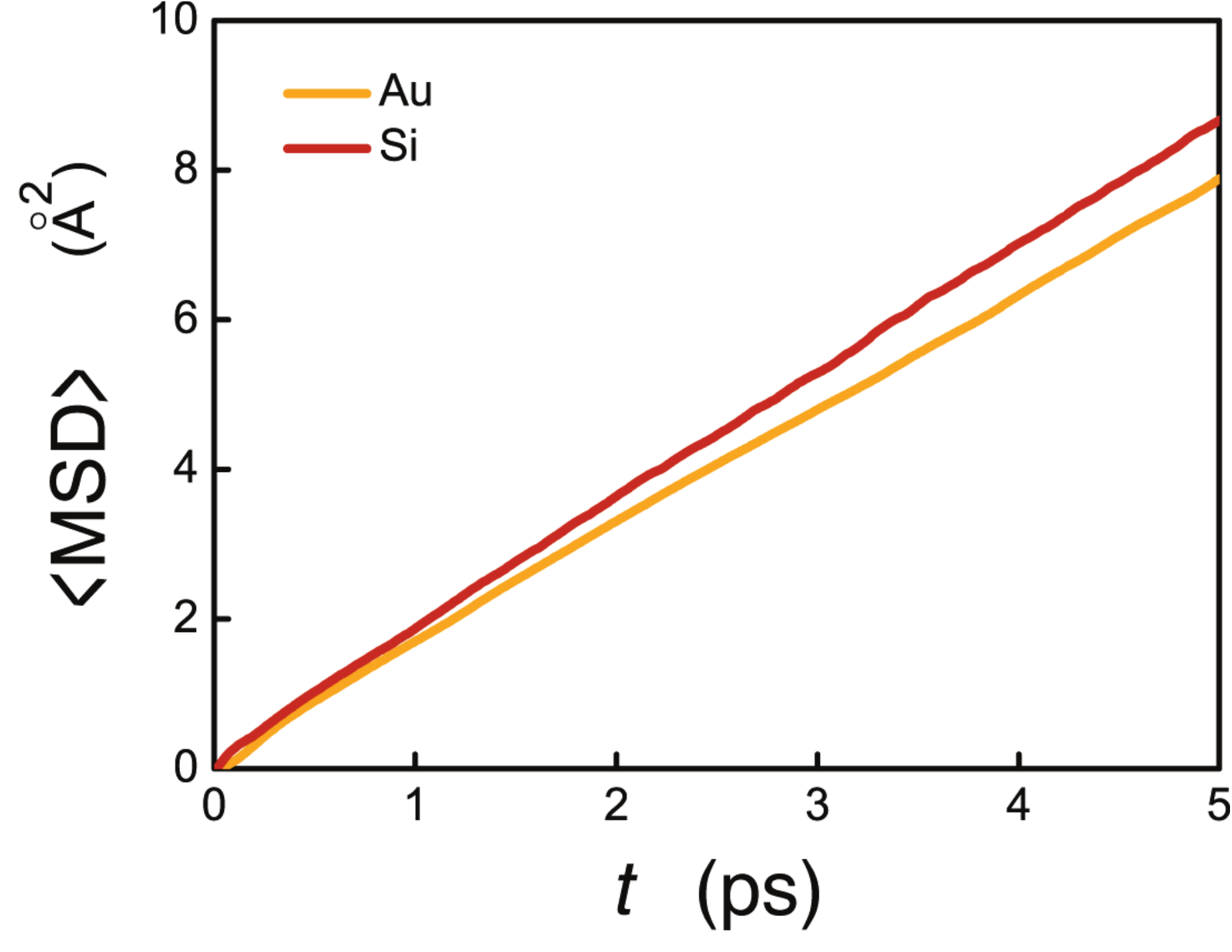}
\end{center}
\caption{Temporal evolution of the ensemble-averaged mean square displacement of surface and bulk atoms in equilibrium MD simulation of an equilibrated $2R=10$\,nm isolated Au$_{53}$Si$_{46}$ particle.
\label{fig:Supp3}
}
\end{figure}

\begin{figure}[h!tb]
\begin{center}
\includegraphics[width=0.8\columnwidth]{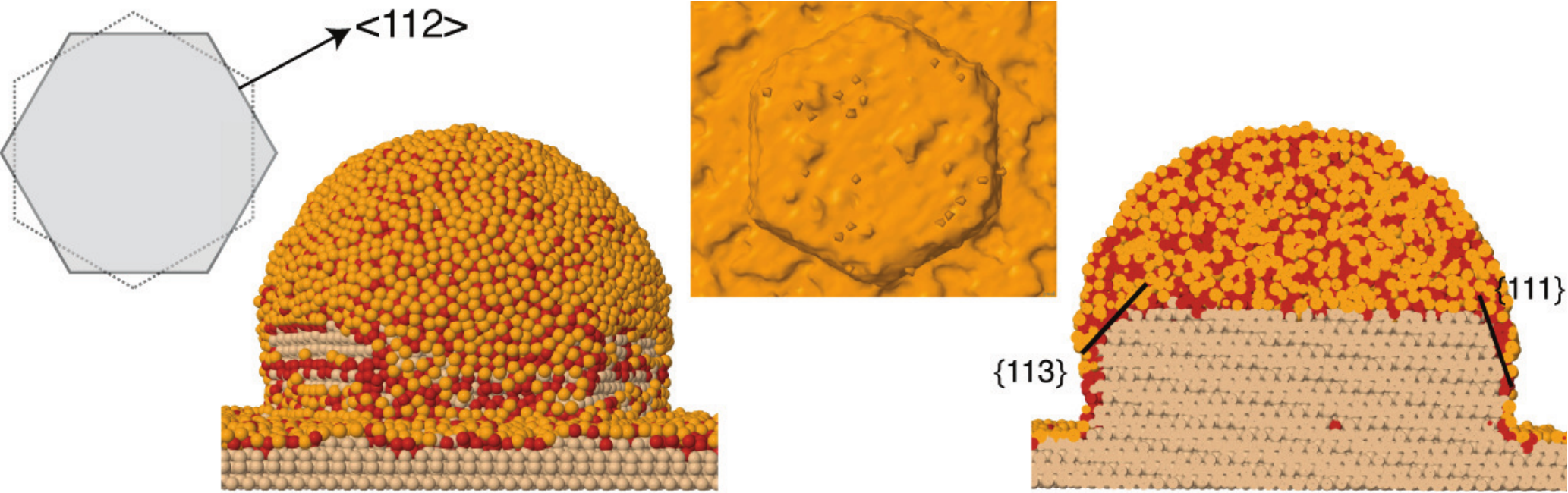}
\end{center}
\caption{Equilibrium morphology of a $2R=10$\,nm nanowire with \{112\} sidewall facets. (left inset) Schematic showing the relationship with 110 sidewall facets. (left) Atomic configuration of the complete nanowire-droplet-substrate system. (middle inset)  (right) A mid-section slice along one of the $\langle$$112$$\rangle$ directions. Dashed lines in the figure indicate the truncating facets.
The SiNW structure is the same as in the double-capped simulations, i.e. it has a hexagonal cross-section with \{110\} facets. The nanowire stem is $3.5$\,nm in length.
\label{fig:Supp4}
}
\end{figure}

\pagebreak

\begin{figure}[h!tb]
\begin{center}
\includegraphics[width=\columnwidth]{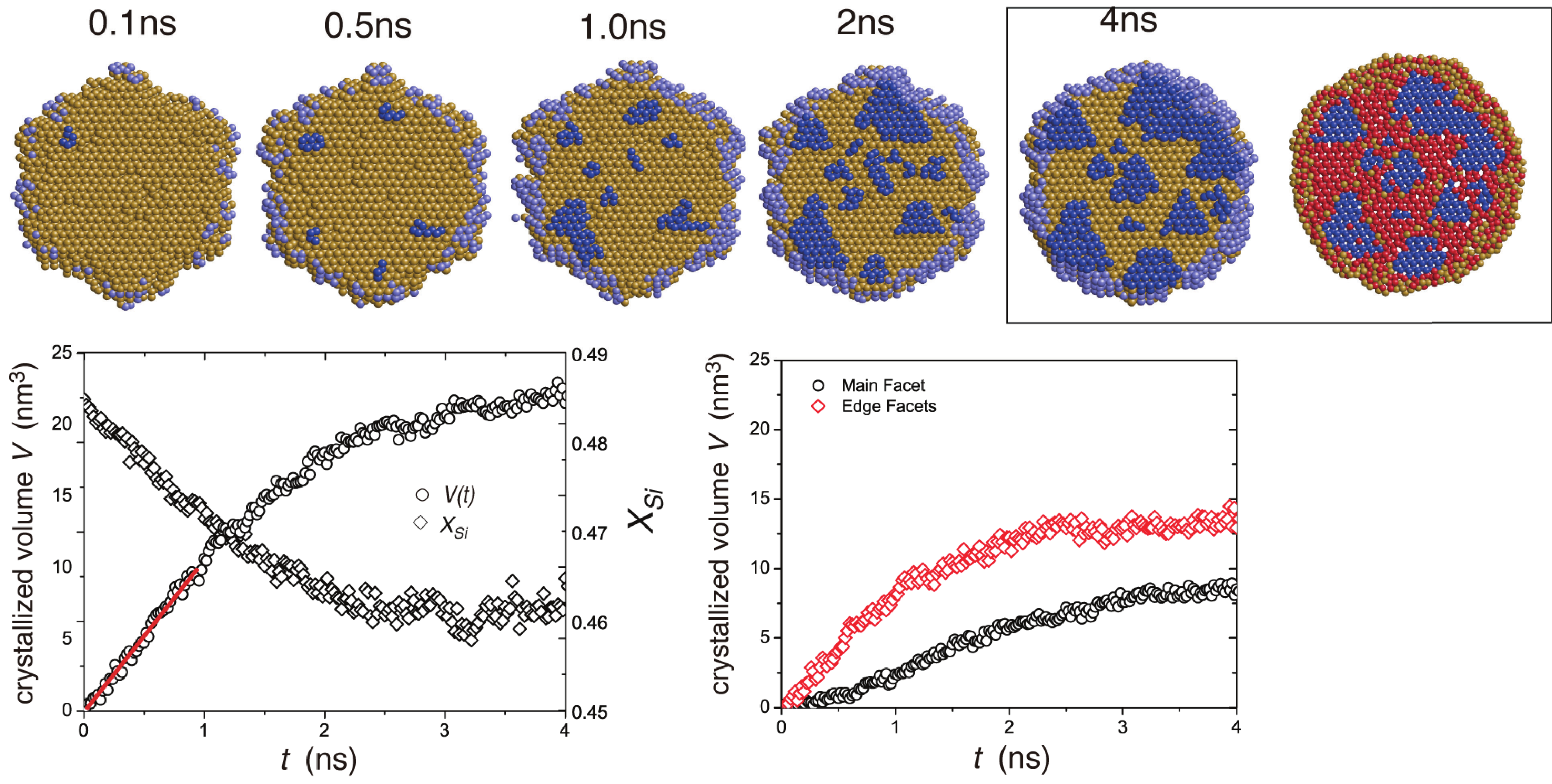}
\end{center}
\caption{Same as in Figure~5 in the main text, but for $X_{Si}=0.485$ ($\Delta_T\approx45$\,K). The $t=4$\,ns configuration is shown with and without liquid Si (red) atoms. We see multiple growing nuclei on the main facet indicating that they are supercritical. The nuclei form preferentially near the corner edges where the main facet meets two truncating facets.}
\label{fig:Supp7}
\end{figure}

\pagebreak

\begin{figure}[h!tb]
\begin{center}
\includegraphics[width=0.7\columnwidth]{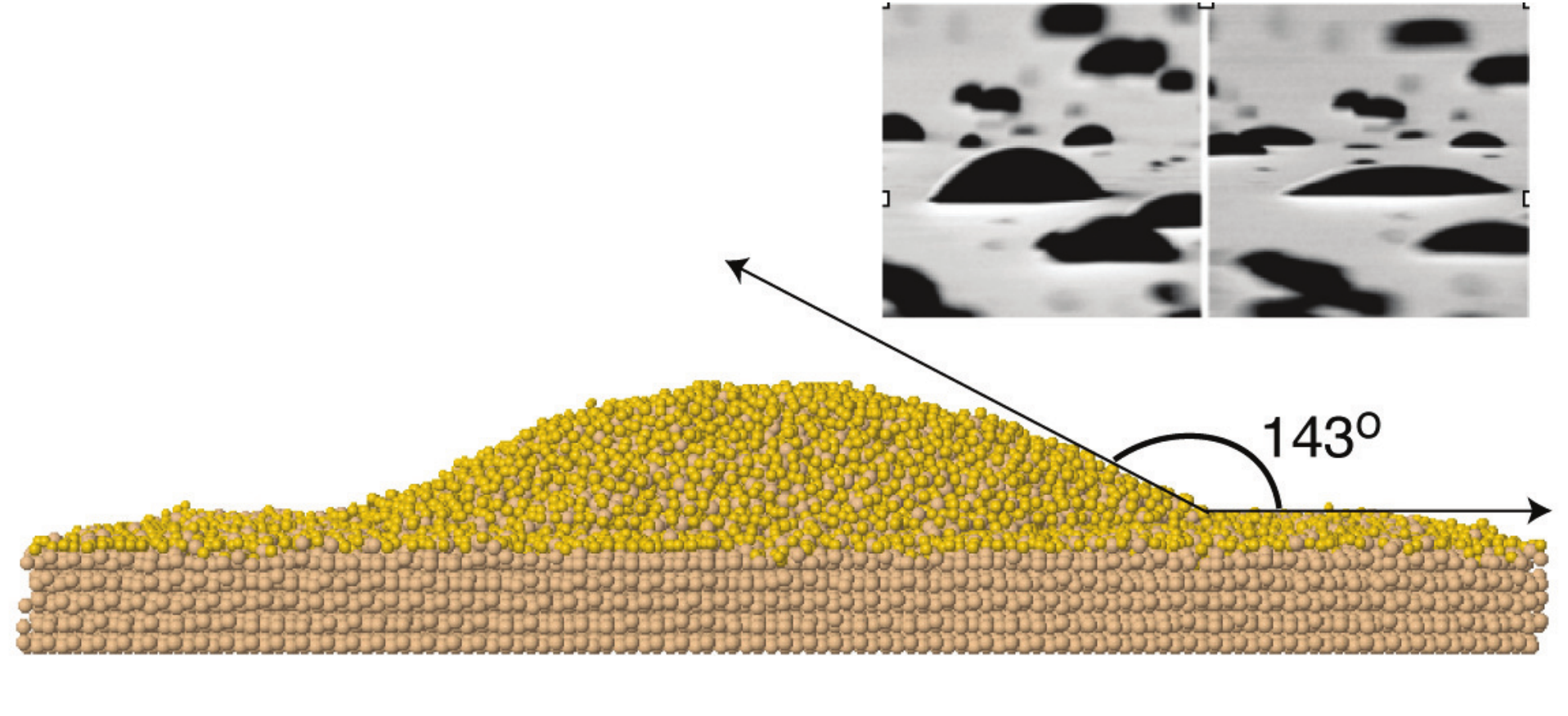}
\end{center}
\caption{Atomic-configuration of an equilbrated eutectic droplet Au$_{33}$Si$_{67}$ on a Si(111) surface. The color scheme is similar to that in the main text with the exception that we make no distinction between liquid and solid Si atoms. The inset shows the morphology of eutectic AuSi droplets on Si(111) reported in prior experiments~\cite{tsf:ResselHomma:2003}. The image is reproduced with the authors' permissions.
\label{fig:Supp0}
}
\end{figure}

\pagebreak
\clearpage

\section*{Supplementary Methods}

\subsection*{Atomic-scale simulations}
The semi-grand canonical Monte-Carlo simulations employed to relax the system at the growth temperature consist of translational and Au$\leftrightarrow$Si exchange (transmutation) moves~\cite{book:AllenTildesley:1989, gbseg:Seidman:2002}. The ratio between translational and exchange moves is fixed at $10:1$. The acceptance ratio for translation moves  is constrained to $50\%$ by adjusting the magnitude of the translations. The equilibration is performed until the energy, pair distribution functions and density profiles converge. In some cases, canonical (NVT) Monte-Carlo computations are performed by turning off the transmutation moves.

The MD simulations are performed by integrating the Newton's equations of motion using velocity-Verlet scheme. The time step is $1$\,fs and the trajectories for the canonical ensemble are generated using a Nos\'e-Hoover to fix the temperature at $873$\,K~\cite{book:AllenTildesley:1989}. Isobaric isothermal simulations (NPT) are performed by using a Parinello-Rahman barostat that allows the cell volume to adjust to the imposed press. The fictitious mass of the barostat is fixed at XX.

\subsection*{Contact angle of  $2R=10\,$nm droplet on Si(111) surface}
A $2R=10$\,nm size spherical droplet consisting of $\sim32000$ atoms is constructed by assigning random coordinates about positions corresponding to average Au-Si equilibrium distance based on the model angular EAM potential. One half of the droplet is then placed on an atomically flat  Si(111) substrate (thickness 8 bilayers). The substrate dimensions are as mentioned in the main text. It is periodic along the in-plane directions. The bottom layer is fixed to prevent the entire system from translating. The exposed regions are decorated with a $(\sqrt{3}\times\sqrt{3})\uppercase{R}\,30^\circ$ Au monolayer and then entire system is then equilibrated to the growth temperature using SGMC minimization followed by NVT MD. Upon equilibration, we see some evidence for disorder within the monolayer. Systematic thin film studies indicate that for our model system, the monolayer undergoes a surface transition at and around $900^\circ$K. A similar transition has been reported in experiments~\cite{tsf:NagaoHenzler:1998}, albeit at $1000^\circ$K. The difference between the experimental and computational contact angles is around $7^\circ$. We attribute this mainly to the differences in the manner in which the droplets attain the eutectic composition. In experiments, the silicon from the substrate melts into the initially Au-rich droplet to arrive at the equilibrium composition, resulting in a the crater like morphology of the substrate. On the other hand, the droplet in the computations is already at liquidus composition before it stabilizes on the substrate. The substrate is almost flat and therefore increases the measured contact angle with respect to the substrate surface plane.



\subsection*{Charge gradient corrections for AEAM}
The computational cell for the MD simulations is chosen to be an FCC crystal slab with dimensions $10\,a_0\times10\,a_0$ in the periodic surface plane, and a thickness of $8a_0$. The slabs are melted and equilibrated for $1$\,ns; the effect of temperature was studied by equilibrating the system for $250$\,ps. Data runs $1$\,ns in duration are carried out for each temperature $T$. The mechanical definition is used to extract the surface tension~\cite{intseg:YuStroud:1997},
\begin{equation}
\tag{S1}
\gamma=\frac{1}{2}\int_{-\infty}^{\infty}[\sigma_{xx}+\sigma_{yy}-2\sigma_{zz}]\le\frac{1}{S}\sum_{i}\{\frac{p_x^2+p_y^2-2p_z^2}{2
m_i}+\sum_{i{\neq}j}\frac{f_xr_x+f_yr_y-2f_zr_z}{4}\}>\,.
\end{equation}
Here the $\sigma_{\alpha\beta}$ are components of the surface stress tensor, $\alpha, \beta=1, 2, 3$ along the $x$, $y$ and $z$ directions in a Cartesian coordinate system. $S$ is the surface area, $m_i$ is the mass of atom $i$, $p_\alpha$ is the ${\alpha}$th momentum component of atom $i$, $r_\alpha$ is the ${\alpha}$th component of the distance atoms $i$ and $j$, $f_\alpha$ is the ${\alpha}$th component force component of atom $i$, and both the summations over $i$ and $j$ run over all the atoms in the systems.

The Au-Au interactions employed as part of the AEAM framework used in this study are of the form suggested in Ref.~\citep{tsf:Zhou:2001}. For pure liquid Au at $T=1400$\,K, the extracted surface tension is $0.9$\,J/m compared to the experimental value of $1.11$\,J/m. The difference is significant and can affect the segregation qualitatively. To test this, we employ a charge-gradient correction~\cite{intpot:WebbGrest:1986} to fit the Au-Au interactions to the experimental value of the liquid/vapor surface tension; the Au-Si and the Si-Si interactions remain unchanged.

In this framework, the potential energy $U_i$ of atom $i$ is,
\begin{align}
& U_i=\sum_i\left\{F(\eta_i)+\sum_{j\neq{i}}\frac{1}{2}\Phi(r)\right\}\nonumber\\
& \eta_i=\rho_i+\beta|\nabla\rho_i|^2,\quad \rho_i=\sum_{j\neq{i}}f(r)\,.
\tag{S2}
\end{align}
Here $F$ is the usual embedding potential energy, $\Phi$ is the pair potential energy, $\rho_i$ the local charge density at atom $i$, and $\eta_i$ is the local charge density of atom $i$ with gradient corrections, and $r$ is the distance between atoms $i$ and $j$. The constant $\beta$ emerges as the main fitting parameter for the liquid surface tension. Notice that the formulation in Ref.~\citep{intpot:WebbGrest:1986} introduces another fitting parameter $c$ which is set to zero here. Then, the $\alpha$th component of force $m_ia_i^\alpha$ on atom $i$ becomes,
\begin{align}
\tag{S3}
m_ia_i^\alpha=&-\sum_{j{\neq}i}\left(\frac{1}{2}\frac{\partial\Phi}{\partial{r}}+\frac{\partial{F}}{\partial{r}}\right)\frac{r_{\alpha}}{r}\nonumber\\
&-2\beta\frac{\partial{F}}{\partial{\eta_i}}
\left\{\sum_{j{\neq}i}\frac{\partial{f}}{\partial{r}}\frac{r_\alpha}{r}\sum_{j{\neq}i}\frac{\partial{f}}{\partial{r}}\frac{1}{r}+
\sum_{j{\neq}i}\frac{\partial{f}}{\partial{r}}\frac{r_\beta}{r}
\sum_{j{\neq}i}\frac{{r_\alpha}{r_\beta}}{r^2}(\frac{\partial^2{f}}{\partial{r^2}}-\frac{\partial{f}}{\partial{r}}\frac{1}{r})\right\}\,,\nonumber
\end{align}
where the first term is the standard EAM-based force and the second term is the charge gradient correction contribution. 
Using this formulation, we have re-calculated the surface tension of pure liquid Au $\gamma$ as a function of the fitting parameter $\beta$ at 1400\,K. The results are shown in Supplementary Figure S2. The surface tension increases monotonically with the magnitude of $\beta$ with the value $\beta=-0.025$ in  excellent agreement with the experimental value. Simple tests are first performed to quantify the effect of this modified potential on the bulk properties of Au at $1400$\,K using MD simulations. For $\beta=-0.025$, the cohesive energy is $-3.626$\,eV/atom and the lattice constant is $a_0=4.197A$. As comparison, for $\beta=0$ which reduces to the original AEAM potential, the corresponding values are $-3.624$\,eV/atom and $a_0=4.206$\,{\AA}, respectively.

This modified Au-Au potential within the AEAM formulation is employed to recalculate the density profiles for AuSi thin films. The surface segregation in AuSi film with $33\%$ Si for charge gradient correction $\beta=0.025$, which is also qualitatively same the result without charge gradient correction. The ensemble-averaged segregation profiles are show in Fig.~\ref{fig:Supp5}B. Again, we see strong subsurface Si segregation which is decorated by a submonolayer of Au. The quantitative differences are minimal, notably that the peak associated with Au submonolayer decoration on the surface is slightly decreased.

\section*{Supplementary Discussion 1}
We estimate the nucleation barrier on the planar Si(111)-AuSi interface by assuming that the nuclei is a circular 2D island. The barrier for homogeneous nucleation on the planar Si(111) surface is roughly $\Delta G \sim \sigma^2h/(\rho\,\mu_{sl})$, where $\sigma$ is the step energy per step step (bilayer) height $h$, $\rho$ is the density of solid silicon, and $\mu_{sl}$ is the free energy difference between solid and liquid phases. Our extracted values of the step enthalpies on Si(111) vary between $\sigma h\approx0.16$\,eV/nm. $\mu_{sl}$ for the small supersaturations in the AuSi model system is $\sim100\,\Delta_X$\,kJ/mol, approximated as the enthalpy of mixing predicted by the model AuSi system~[34]. Using these values yields a barrier $\Delta G\approx20$\,meV$/\Delta_X$ that is of the same order as the extracted value of $\Delta G\approx7$\,meV$/\Delta_X$. The self-consistent comparison is strong evidence that the nucleation barrier during nanowire growth is lowered due to the multi-faceted morphology of the growth front.

\section*{Supplementary Discussion 2}
The Si incorporation rate follows from the volume growth rate due to the flux made available by the surface catalysis. It scales as $k_{vl}\,p\,A_s$, where $p$ is the precursor gas partial pressure, $k_{vl}$ is normalized catalyst rate constant per unit partial pressure, and $A_s$ is the exposed surface area of the droplet. The rate constant reported in recent experiments on growth of Si nanocrystals within an AuSi droplet exposed to a silane flux is $k_{vl}\sim10^3\,\, {\rm nm\, s}^{-1}\,{\rm Torr}^{-1}$~[26]. Approximating the droplet surface area to that of a hemisphere of radius $5$\,nm, under typical processing conditions ($p\approx10^{-5}$ Torr) yields the incorporation rate assumed in the text. As a simple check, the axial nanowire growth rate then is of the order of a nanometer per minute which is consistent with growth rate observed in in-situ experiments.

The transition from flux-controlled to nucleation-controlled growth occurs when the flux due to catalysis equals the crystallization rate on the main facet,
\begin{equation}
\tag{S4}
v=v_0\exp\left(-\frac{\Delta G/\Delta_x}{k_BT}\right)\nonumber
\end{equation}
From Fig~6b in the main text, we have $v_0=0.58$\,m/s and $\Delta G/k_BT=0.055$. For Si volume flux of the order of $I_{Si}\approx1$\,atom/ms, the axial nanowire growth rate is,
\begin{equation}
\tag{S5}
v=\frac{I_{Si}}{\rho\pi R^2}\nonumber,
\end{equation}
where $\rho=50$\, atom/nm$^3$  is the density of crystallized Si and $R=5$\,nm is the nanowire radius in our simulation.
Equating Eqs.~S1 and S2, we get $\Delta_x=0.002$.

\end{widetext}

\end{document}